\documentclass[journal]{IEEEtran}
\ifCLASSINFOpdf
\else
\fi

\usepackage[T1]{fontenc}
\usepackage[utf8]{inputenc}
\usepackage{graphicx}
\usepackage{hyperref}
\usepackage{multirow}
\usepackage{tabularx}
\usepackage{color}
\usepackage{textcomp}
\usepackage{tipa}
\usepackage{amsmath}
\usepackage{amssymb}
\usepackage{amsfonts}
\usepackage{amsxtra}
\usepackage{wasysym}
\usepackage{isomath}
\usepackage{mathtools}
\usepackage{txfonts}
\usepackage{upgreek}
\usepackage{enumerate}
\usepackage{tensor}
\usepackage{pifont}
\usepackage{soul}
\usepackage{arydshln}
\usepackage[caption=false,font=normalsize,labelfont=sf,textfont=sf]{subfig}
\usepackage{makecell}
\usepackage{cite}

\begin{document}
\title{A Study on mm-wave Propagation in and around Buildings}

\author{Leonardo~Possenti,
        Marina~Barbiroli,
        Enrico~M.~Vitucci,
        Franco~Fuschini,
        Mattia~Fosci,
        and~Vittorio~Degli-Esposti% <-this % stops a space
\thanks{This work was funded in part by the Italian Ministry of University and Research (MUR) through the programme ”Dipartimenti di Eccellenza (2018–2022) — Precision Cyberphysical Systems Project (P-CPS),” and in part by the Eu COST Action INTERACT (Intelligence-Enabling Radio Communications for Seamless Inclusive Interactions) under Grant CA20120.}
\thanks{L. Possenti, M. Barbiroli, E. M. Vitucci, F. Fuschini, and V. Degli-Esposti are with the Dept. of Electrical, Electronic and Information Engineering "G. Marconi", CNIT, University of Bologna, 40136 Bologna, Italy (e-mail: leonardo.possenti2, marina.barbiroli, enricomaria.vitucci, franco.fuschini, v.degliesposti @unibo.it ).}% <-this % stops a space
\thanks{M. Fosci is with JMA Italy/TEKO TELECOM s.r.l., 40024 Castel San Pietro Terme, Bologna, Italy.}% <-this % stops a space
}




\maketitle

\begin{abstract}
mm-waves are envisaged as key enabler for 5G and 6G wireless communications, thanks to the wider bandwidth and to the possibility of implementing large-scale antenna arrays and new advanced transmission techniques, such as massive MIMO and beamforming, that can take advantage of the multidimensional properties of the wireless channel. In order to further study the mm-wave wireless channel, where propagation shows different characteristics compared to the sub-6 GHz band, a joint measurement and simulation campaigns in indoor and outdoor microcellular environments has been carried out. The investigation highlights that the traditional assumption that mm-wave NLoS propagation is problematic is not true since significant reflections, scattering and even transmission mechanisms provide good NLoS coverage in most indoor and outdoor scenarios. This also reflects in the limited angle-spread differences between LoS and NLoS locations in some cases. Finally, the contribution of different propagation mechanisms (reflection, diffraction, scattering and combination of them) to the received power is analyzed in the paper with the help of ray tracing simulations.
\end{abstract}


% \fi
\IEEEpeerreviewmaketitle

\section{Introduction}
\IEEEPARstart{I}{n} order to cope with the explosive growth of data rate demand and given the sub\hbox{-}6 GHz spectrum shortage, millimeter wave (mm-wave) frequencies (30-300 GHz) have been considered for 5G systems and beyond \cite{1,2,3}. The primary motivation for using mm-wave spectrum is its ability to provide extremely high data rates and low latencies due to the large available bandwidth, enabling services with very stringent requirements, such as wireless cognition, centimeter-level location, and ultra-high-definition video and audio streaming. Moreover, higher frequencies allow for the use of smaller antenna elements, and therefore of massive antenna arrays and advanced transmission techniques, e.g., massive MIMO (mMIMO) and pencil-beamforming/beamtracking that are envisioned to be key enablers for future systems \cite{1}. Furthermore, massive arrays, due the high-gain beams, can compensate for the very high isotropic attenuation at these frequencies, and also provide increased capacity in high-density, multiple users scenarios (MU-MIMO) \cite{4}.

The effectiveness of multiantenna techniques rely on the multidimensional and spatial properties of the wireless channel, as these technologies necessitate special strategies \cite{5} for beam acquisition and tracking, which requires deep knowledge of the mm-wave wireless channel. Therefore, the design of wireless systems operating in the mm-wave band will benefit from a thorough multipath propagation characterization.

Extensive measurement campaign analysis \cite{6,7,8,9} have shown that high-frequency communications face limitations in terms of propagation mechanisms, compared to the sub-6 GHz band. Diffractions around, and transmissions through, obstacles are less prominent at mm-wave frequencies, while dominant paths, apart from the line-of-sight (LoS) one, only include a few reflected or scattered paths. Few studies have tried to analyze the inner nature of the mm-wave channel, i.e. what propagation processes actually generate such channel characteristics \cite{10,11,12,13}, but some important issues related to the actual reverberation degree of mm-wave indoor propagation, the relative importance of multiple-bounce vs. single-bounce reflections, of diffraction vs. diffuse scattering have not been completely addressed so far.

Indeed, such issues are important for the development of system level simulators or fast and reliable ray-based, spatially consistent propagation models that are needed for the design, deployment and functioning of the coming wireless systems. In particular, recent research advocates for the real-time use of deterministic ray-based propagation models to estimate the channel state and cope with abrupt channel changes (e.g., blockage) or to help initial beam acquisition and tracking.

Therefore, beyond the validation of the Ray Tracing (RT) simulation tool at mm-waves, this paper aims at filling the gaps in the comprehension/identification of the relevant propagation mechanisms at mm-waves, in order to effectively design/plan future wireless network. Directional measurements and RT simulations of indoor and outdoor short-range propagation environments at 27 GHz and 38 GHz are combined to investigate channel properties, such as angular dispersion in LoS and NLoS conditions, multipath richness in indoor environment, the relative importance of multiple-bounce vs. single-bounce reflections, of diffraction vs. diffuse scattering and the relevance of through-wall transmission, taking into account the different environment characteristics.

The considered scenarios are typical indoor and outdoor environments in the proximity of buildings, where a high density of users, with different requirements in terms of QoS, is expected. All the environments show large spaces with a significant number of scattered elements (columns, stairs, doors, plants, walls of different materials) and therefore are quite suitable for the characterization of mm-wave multipath propagation.

The paper is organized as follows: in Section {\hyperref[sec:2]{II}} measurements are firstly described with the measurements set-up, then in Section {\hyperref[sec:3]{III}} the RT simulations are presented, while in Section {\hyperref[sec:4]{IV}} results of the measurements campaign and the simulation outputs are shown, finally Section {\hyperref[sec:5]{V}} includes the main conclusions of this work.

\section{Measurement Campaigns}\label{sec:2}

Directional measurements at 27 and 38 GHz have been first performed in a large\hbox{-}indoor environment in the hall of the School of Engineering of the University of Bologna, then in an outdoor setting in the courtyard of the School of Engineering, using a portable spectrum analyzer and a rotating positioner with directive antennas over multiple receiving points.

Besides, in order to increase the environment diversity, further measurements (non-directional) have been carried out at the frequency of 28 GHz in a typical open-plan indoor office environment located in the main headquarter of the company TEKO-JMA Wireless, Castel S. Pietro, Italy.

\subsection{University measurement setup}

The equipment used to sound the radio channel is a portable Spectrum Compact Analyzer (SCA) by SAF Tehnika \cite{14}, composed of a spectrum analyzer, a continuous wave signal generator and two horn antennas, along with cables, connectors and two tripods. The SCA is a very light, battery-powered analyzer, which is an attractive solution because of its portability and easy use.

The signal analyzer is connected through an USB cable to the laptop to store the information of the received signal using a dedicated proprietary software GUI. The signal generator is instead connected to the transmitting antenna. The directive horn antennas operate in a frequency band ranging from 26.5~GHz to 40.5 GHz and the considered frequencies are the 27~GHz and the 38 GHz, as they are two frequencies of interest for 5G. Technical data are reported in Table I.

Through a rotating antenna positioner, the receiver is steered to scan the channel from each angular direction, in the azimuth plane (0$^{\circ}$-360$^{\circ}$), with a step of 15$^{\circ}$, for a total of 24 directions.

\begin{table}[!ht]
\centering
\caption{University Measurement Setup}
\begin{tabularx}{0.49\textwidth}{ c c c c }
 \hline
 \hline
 Frequency & Antenna gain [dBi] & Antenna HPBW & \makecell{TX Output\\Power [dBm]}\\
 \hline
 27 GHz & 20.5 & \makecell{E-plane: 14$^\circ$\\H-plane: 17.5$^\circ$} & 5\\
 \hline
 38 GHz & 21.5 & \makecell{E-plane: 11.5$^\circ$\\H-plane: 13$^\circ$} & 5\\
 \hline
\end{tabularx}
\end{table}

\subsection{University measurement campaign}

\begin{figure}[!ht]
\centering
\includegraphics[width=0.45\textwidth]{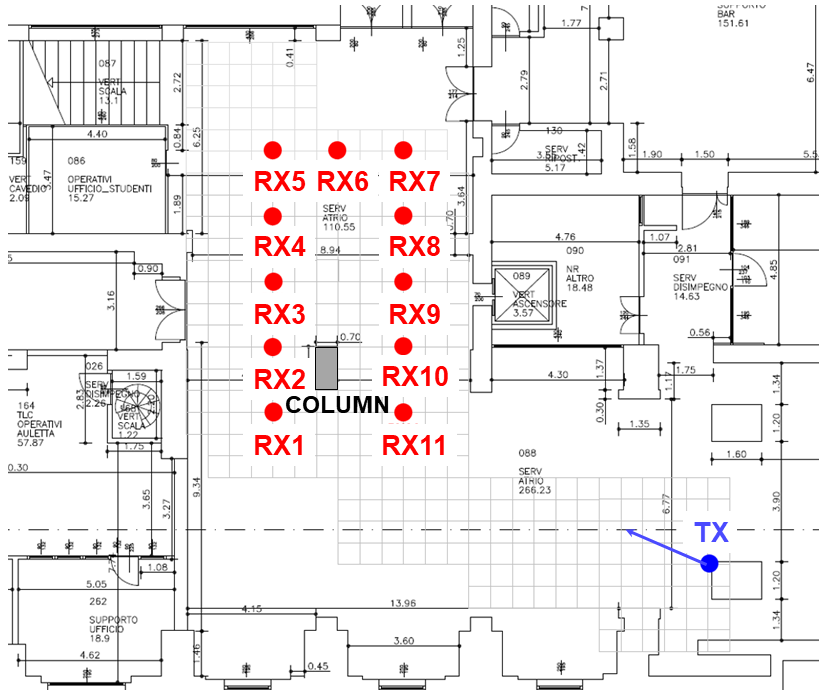}
\caption{Map of the entrance hall of the School of Engineering of Bologna University.}
\label{fig:1}
\end{figure}

Indoor channel measurements are conducted in the entrance hall of the University of Bologna (UniBO), reported in {\hyperref[fig:1]{Fig 1}}. The environment is an open space, almost empty, with sparse corners and a central column, made of travertine. The floor is entirely made of a particular type of marble, while the walls are composed by travertine. In the top side of the figure, there is a large, windowed glass wall. Ceiling is at a height of 5-6 meters.

The measurement arrangement is sketched in {\hyperref[fig:1]{Fig 1}}, the TX unit is located at the bottom-right corner of the figure (blue dot), whereas the RX unit is placed in 11 different central locations in the room (red dots), both in LoS (RX1, RX3, RX10, RX11) and in NLoS locations. The height of TX and RX antennas are the same for all positions (2.1 m). The TX is pointing towards the direction of the RXs (blue arrow in {\hyperref[fig:1]{Fig 1}}), so as for the RXs to fall inside the main lobe of the TX antenna. The environment is kept almost static during the measurement.

\begin{figure}[!ht]
\centering
\includegraphics[width=0.49\textwidth]{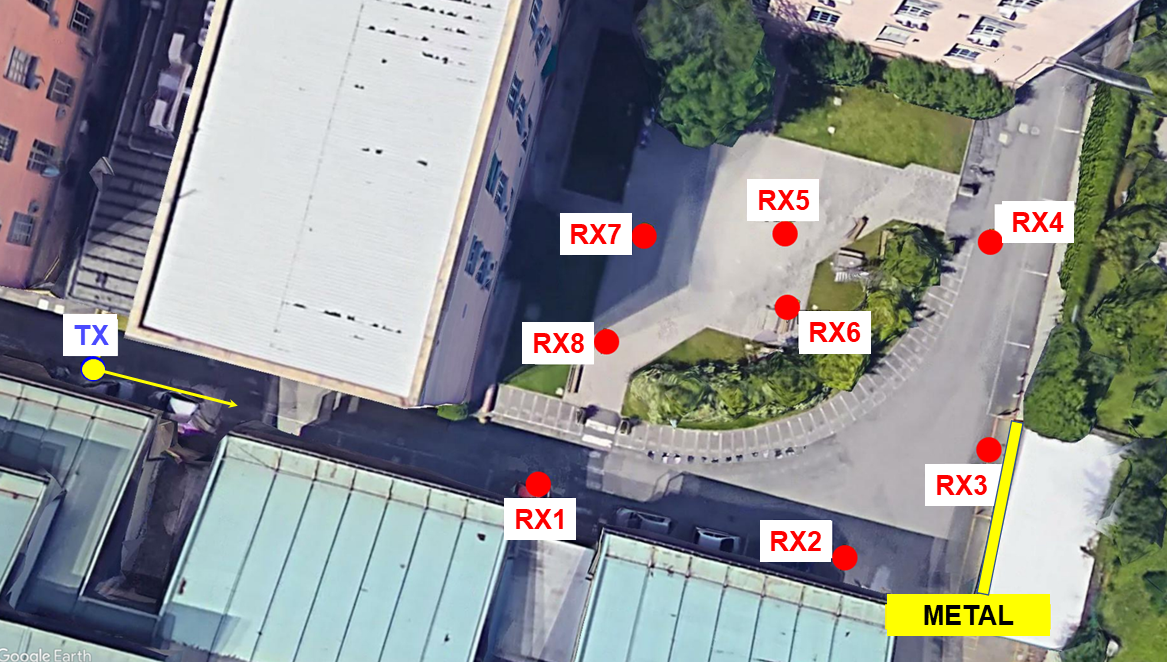}
\caption{Outdoor courtyard in the School of engineering of Bologna University.}
\label{fig:2}
\end{figure}

Outdoor channel measurements are conducted in the internal yard, around buildings of the University of Bologna, reported in {\hyperref[fig:2]{Fig 2}}. The environment has an open garden surrounded by high buildings and a short canyon-street, where the transmitter is placed. The TX is pointing along the small street canyon as shown in {\hyperref[fig:2]{Fig 2}} (yellow arrow).

The construction at the bottom-right corner is made of metal. The RX unit is placed in 8 different locations both in LoS (RX1 and RX2) and in NLoS. TX and RX antennas were kept at the same height for all positions (2.1 m).

\subsection{JMA measurement set-up}

The measurements in the indoor office environment have been carried out using a 5G New Radio (NR) transceiver, which generates an OFDM signal at the frequency of 27.925 GHz (N258), with a bandwidth of 100 MHz. The average power on the bandwidth was considered, by sounding the so-called Reference Signal Received Power (RSRP). In particular, the received power has been sounded with a portable Real-Time Spectrum Analyzer (VIAVI Celladvisor 5G), which was able to demodulate the NR signal and to extract the Synchronization Signal RSRP (SS-RSRP).

The TX antenna is a 16-elements planar array with 12 dBi maximum gain, half-power beamwidth of 60$^{\circ}$ in the horizontal plane and circular polarization, which was kept with a fixed orientation ({\hyperref[fig:3]{Fig 3}}), while the receiving antenna was an omnidirectional vertically polarized antenna, with 3 dBi maximum gain. The measurement set-up allows for all the receiver locations in the ``Meeting Room'' to be within the main lobe of the TX antenna (strong LOS condition), while the receivers in the ``Guest Room'' are outside the main lobe of the TX antenna and in NLOS condition, then allowing to better analyse the effect of multipath on the received power. For this scenario, only narrowband results are presented in the following due to the different receiving equipment, employing an omni antenna instead of a directive rotating antenna.

\subsection{JMA measurement campaign}

The measurement scenario at the JMA wireless company is an office environment, which consists in an open space with a meeting room and a small guest room located in its center and separated each other and from the outside through partition walls made by plasterboard (see {\hyperref[fig:3]{Fig 3}} (a)). The TX equipment is located at the bottom of the meeting room, and the TX array is pointed perpendicularly in the direction of the opposite wall, as indicated by the black arrow in {\hyperref[fig:3]{Fig 3}} (a). The receiving equipment has been moved in different locations inside the meeting room (red points, 14 locations), the guest room (blue points, 12 locations), and the corridor in front of them (green points, 5 locations). Moreover, additional measurements have been done along an ``external'' receiving route as shown in {\hyperref[fig:3]{Fig 3}} (b), where the receiver has been moved all around the 2 rooms, in the external open space and in the adjacent storeroom located on the right side of the map.

\section{RT Simulations}\label{sec:3}
In order to interpret measurement results, validate the RT model and analyse the main propagation mechanisms taking place in the different environments, RT simulations have been carried out as explained below.

\subsection{UniBO RT tool}

The RT model developed in house at the University of Bologna, also called 3DScat in the following, is an image-based 3D RT tool \cite{15,16}, capable of simulating multipath propagation in indoor and outdoor environments with multiple interactions, including specular reflection, transmission, diffraction, diffuse scattering and any combination of these. In particular, diffuse scattering is simulated here using the Effective Roughness (ER) heuristic model in order to take into account non-specular scattering due to environment details (surface roughness, decorations, smaller objects) that are not described in the environment database.

The objective of such RT simulations is two-fold.

\begin{figure}[!ht]
\centering
\includegraphics[width=0.47\textwidth]{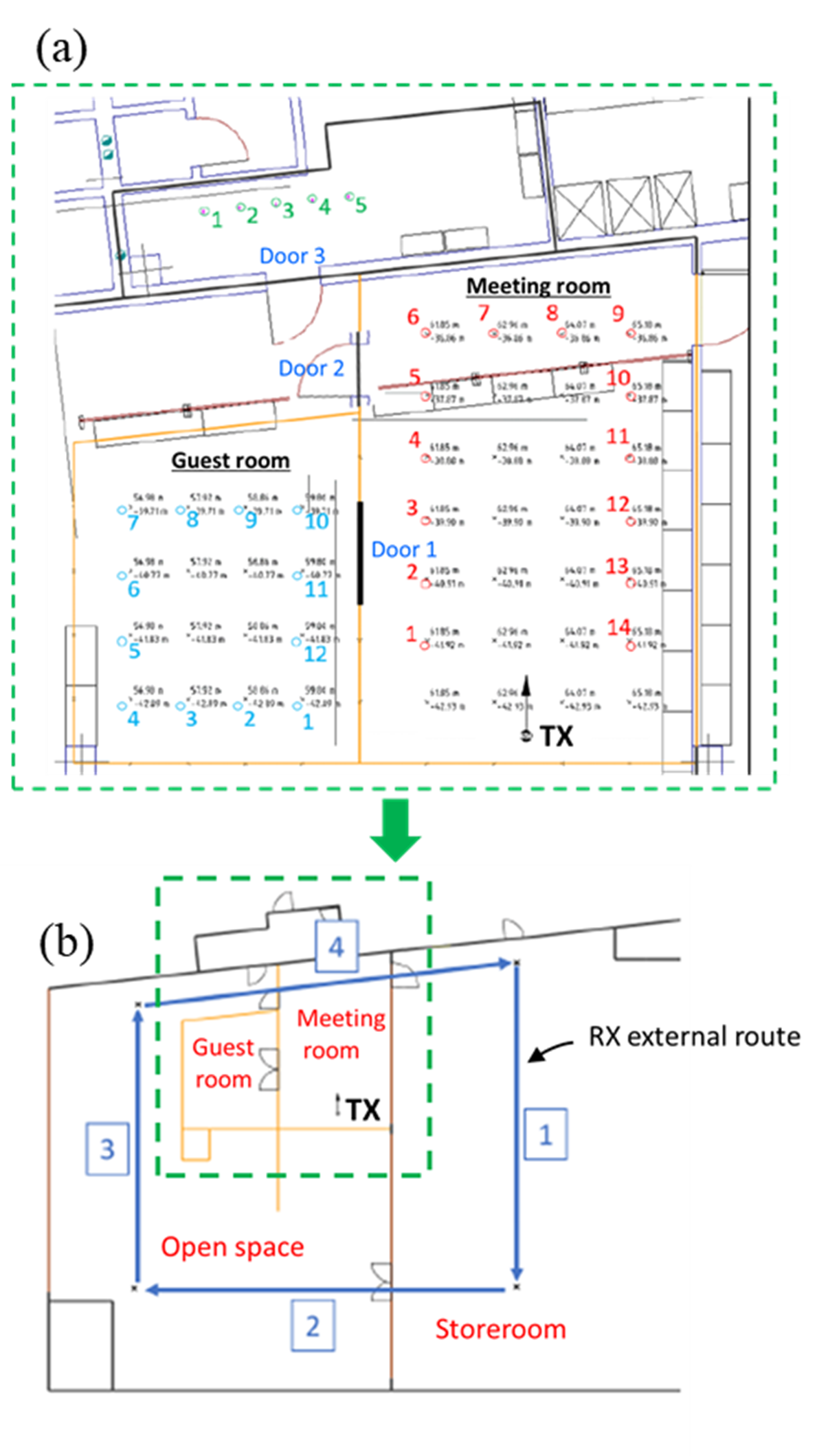}
\caption{Map of the office measurement scenario at JMA Wireless, with TX location, internal (a) and external (b) RX locations.}
\label{fig:3}
\end{figure}

The first one is to calibrate and validate the tool against the mm-wave measurements. To support reliable channel predictions the electromagnetic parameters of the different walls and environment objects for the two frequencies have been derived from literature surveys \cite{17,18},(e.g. values for the plasterboard in the JMA Wireless scenario, {\hyperref[fig:3]{Fig 3}}) and through the Fabry-P\'{e}rot (FP) method presented in \cite{19} (e.g. values of the marble floor in the University Hall, scenario in {\hyperref[fig:1]{Fig 1}}). The FP method can be adopted for an item-level investigation of the e.m. properties of construction materials (i.e., the complex relative permittivity), in order to calibrate the RT tool and get consistent results. When the FP method cannot be applied or does not provide reliable results, and when the considered material is not present in the literature, it is also possible to perform simple reflectivity and transmissivity measurements to manually tune both the real and the imaginary part of the complex permittivity and obtain the most suitable values.

The second significant objective is to use RT to investigate and interpret the measurement results by analyzing the propagation mechanisms (e.g., number and types of bounces) corresponding to each contribution that shows-up as a measured, angle-related contribution. The analysis of the multidimensional propagation mechanisms and the identification of the dominant ones as a function of the environment and the considered frequency is fundamental to correctly plan mm-waves networks. As it will be shown in the following, the analysis performed through the RT tool confirms that at mm-wave frequencies the propagation mechanisms play a different role compared to the sub-6 GHz band, and this has to be taken into account in the planning of future wireless networks, in order to guarantee a sufficient Quality of Service (QoS).

An additional goal is to evaluate RT as a real-time prediction channel tool to assist beamforming techniques (RT-assisted beamforming \cite{20}).

The fundamental structures of the room, such as main walls, floor and ceiling, metal rods and windows, are considered in RT simulations. The main RT-simulation parameters are summarized in Table II: the choice of the maximum number of enabled interactions is a trade-off between computational effort and prediction accuracy.

\begin{table}[!ht]
\caption{UniBO RT simulation settings}
\begin{tabularx}{0.49\textwidth}{
p{\dimexpr 0.79\linewidth-2\tabcolsep-2\arrayrulewidth}
p{\dimexpr 0.21\linewidth-2\tabcolsep-\arrayrulewidth} } \hline
\hline
\centering\arraybackslash{}Maximum number of interactions for each ray & \centering\arraybackslash{}7 \\
\centering\arraybackslash{}Maximum number of reflections for each ray & \centering\arraybackslash{}5 \\
\centering\arraybackslash{}Maximum number of diffractions for each ray & \centering\arraybackslash{}2 \\
\centering\arraybackslash{}Maximum number of transmissions for each ray & \centering\arraybackslash{}2 \\
\centering\arraybackslash{}Maximum number of scatterings for each ray & \centering\arraybackslash{}1\\
\centering\arraybackslash{}Combined reflections and diffractions & \centering\arraybackslash{}Yes (max 3) \\
\centering\arraybackslash{}Combined scattering and reflections & \centering\arraybackslash{}Yes (max 1) \\ \hline
\end{tabularx}
\end{table}

\subsection{iBwave RT tool}

For the indoor office scenario at TEKO-JMA, the RT simulations are also compared with additional simulations performed using the iBwave tool \cite{21}, a commercial suite currently used by the TEKO-JMA company, acknowledged as a reference tool for the planning of wireless networks. iBwave includes a ``simplified'' RT engine which is able to simulate, in addition to the direct (LoS) path, transmission through walls and single-bounce specular reflection.

\section{Results}\label{sec:4}

In this section, the indoor and outdoor measurements at the two considered frequencies are analyzed and compared with RT simulations to investigate the channel propagation characteristics and RT performances. Once validated vs. measurements in the received power and angle domains, the RT tool has been used to perform an analysis of the underlying propagation mechanisms - not possible using measurements only \textendash{} as discussed in section IV.C.

\subsection{Narrowband analysis}

As a first step, the received power and the Path Loss (PL) extracted from measurements in the different scenarios are compared with RT simulations. Regarding the directional measurements carried out in the University scenarios,  measured and RT-simulated received power values are compared for all the 24 Rx antenna azimuth orientations. In Table III simulation performance in terms of RMSE is reported for all the scenarios considered in this study.

\begin{table}[!ht]
\caption{RMSE between measurement and simulations in indoor and outdoor environment}
\begin{tabularx}{0.49\textwidth}{
p{\dimexpr 0.72\linewidth-2\tabcolsep-2\arrayrulewidth}
p{\dimexpr 0.28\linewidth-2\tabcolsep-\arrayrulewidth} } \hline
\hline
\centering\arraybackslash{}ENVIRONMENT & \centering\arraybackslash{}RMSE [dB] \\\hline
\centering\arraybackslash{}Indoor UniBO hall @ 27 GHz (UniBO RT) & \centering\arraybackslash{}1.37 \\
\centering\arraybackslash{}Indoor UniBO hall @38 GHz (UniBO RT) & \centering\arraybackslash{}1.54 \\
\centering\arraybackslash{}Indoor JMA @ 28 GHz (UniBO RT) \par Indoor JMA @ 28 GHz (iBwave) & \centering\arraybackslash{}5.2 \par 6.1 \\
\centering\arraybackslash{}Outdoor UniBO @ 27 GHz (UniBO RT) & \centering\arraybackslash{}4.7 \\
\centering\arraybackslash{}Outdoor UniBO @38 GHz (UniBO RT) & \centering\arraybackslash{}3.6 \\\hline
\end{tabularx}
\end{table}

\begin{figure}[!ht]
\centering
\includegraphics[width=0.47\textwidth]{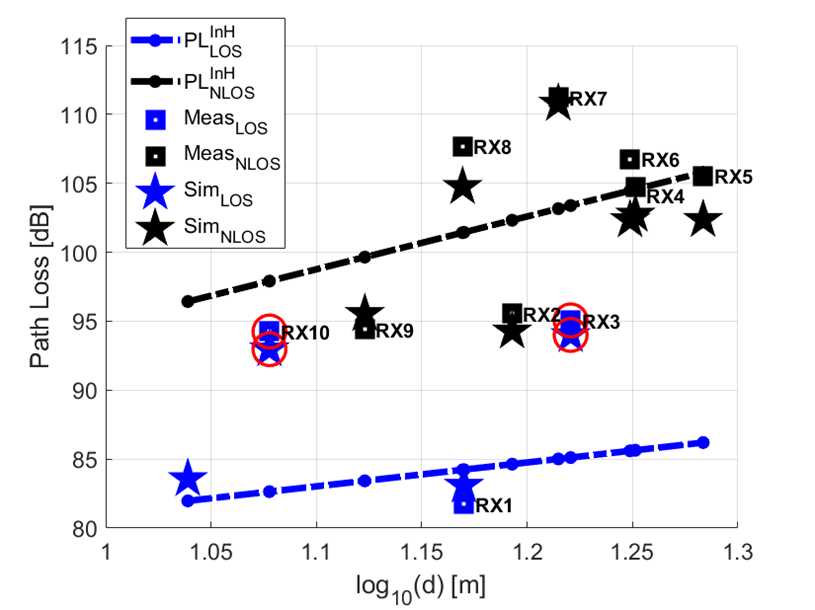}
\caption{Measured and simulated PL, and 3GPP InH model, as a function of the distance, for the indoor UniBO environment @ 38 GHz.}
\label{fig:4}
\end{figure}

The good performance of the RT model is evident, especially in the simple, empty and large environment of the entrance hall of the Engineering School.
In {\hyperref[fig:4]{Fig. 4}}, the 38 GHz local-average PL for each RX location of the indoor hall scenario is reported as a function of the distance, for both measurements (square markers) and RT simulations (star markers). LoS and NLoS receivers are represented in the figure with different colors, blue and black, respectively.

Measured/simulated PL is also compared with the reference propagation models proposed by ITU-T and 3GPP for IMT-2020 system simulations \cite{22}. Specifically, PL values have been compared with the Indoor Hotspot model (InH) described in \cite{22}, for both the LoS (blue dashed line) and NLoS (black dashed line) cases. The InH scenario is intended to take into account various typical indoor deployment cases including offices, open areas, corridors and shopping malls with transmitters at a height of 2-3 m and receivers are at a height of 1.5 m.

As can be noted from {\hyperref[fig:4]{Fig. 4}} there are some Rx locations in a quasi-LoS condition (RX3 and RX10), meaning that they are directly visible from the TX but the Fresnel ellipsoid is partly obstructed. Such receivers (highlighted by the red circle marker) have a very different measured PL compared to what predicted by 3GPP model, being their PL is close to the one of NLoS locations. This behavior is probably due to the low contribution of diffraction and to the critical, sharp transition on the visibility boundary (as reported in Section {\hyperref[sec:4]{IV}}.{\hyperref[sec:4C]{C}}) at mm-waves. A similar trend has been found at 27 GHz, highlighting that a deterministic approach based on an accurate description of the environment is more suitable at mm-wave frequencies, where the traditions LoS/NLoS classification can lead to large errors.

\begin{figure}[!ht]
\centering
\subfloat[]{\includegraphics[width=0.47\textwidth]{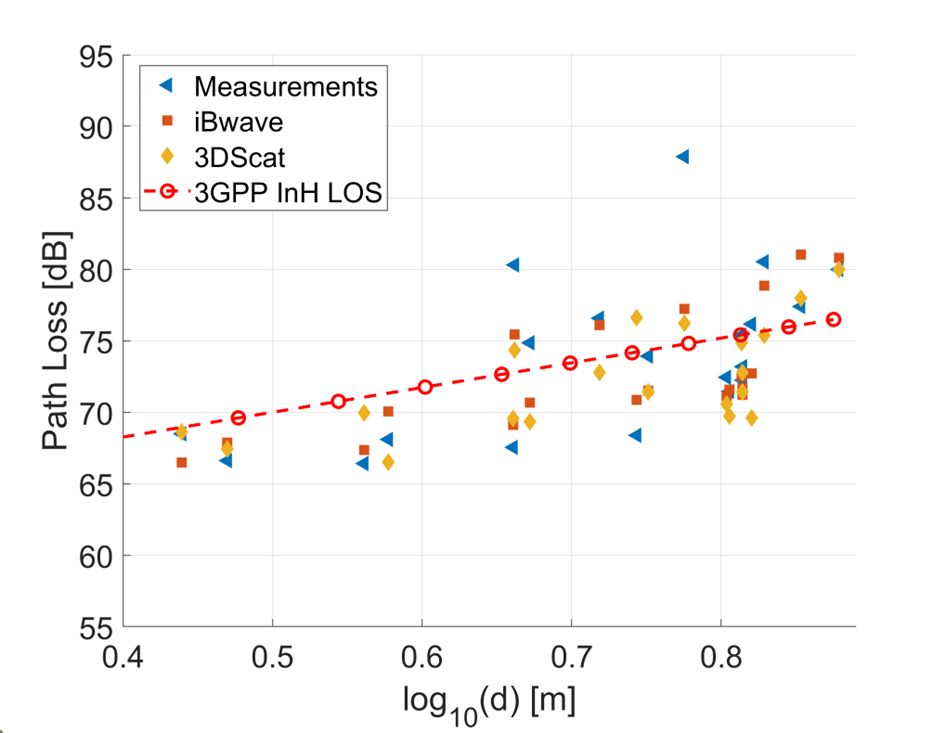}}
\hfill
\subfloat[]{\includegraphics[width=0.47\textwidth]{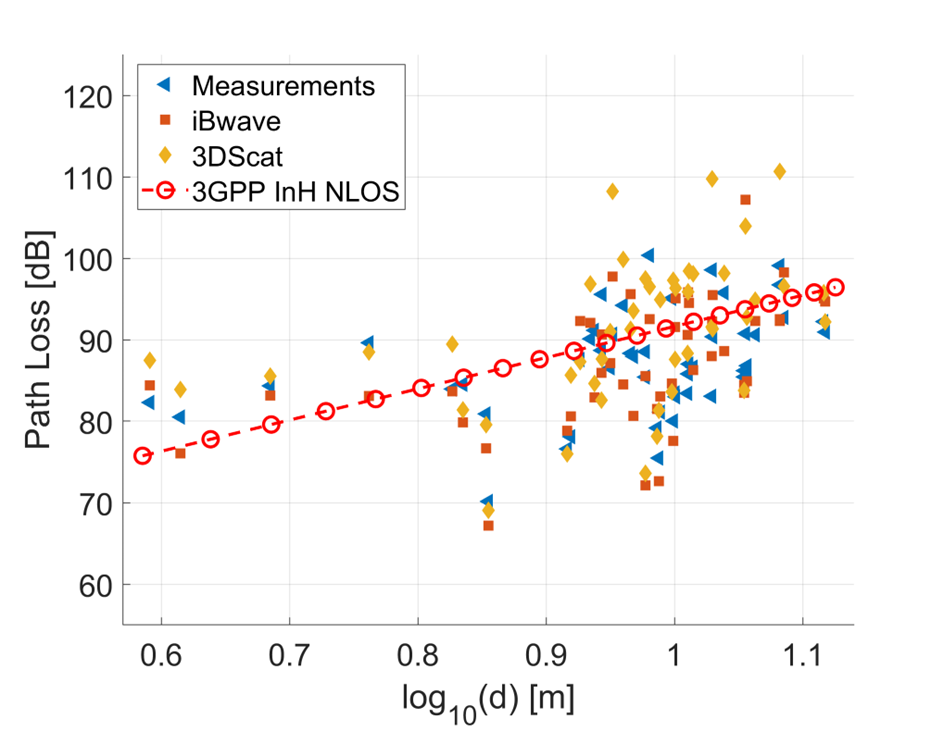}}
\caption{Measured PL and simulated PL (using both the iBwave and the 3DScat tools) and 3GPP InH model as a function of the distance for the indoor JMA environment @ 28 GHz: (a) LoS or quasi-LOS cases; (b) NLOS cases.}
\label{fig:5}
\end{figure}

In {\hyperref[fig:5]{Fig 5}} similar results are shown for the indoor office (JMA Wireless, {\hyperref[fig:3]{Fig 3}}) environment at 28 GHz.

In the plots, measurements are represented with blue triangles, while RT simulations have been carried out using both the UniBO RT tool (orange diamonds) and the commercial iBwave tool (red squares), moreover the red dotted line reports the simulation values for the 3GPP InH model. Results are shown separately for the LoS/quasi-LOS locations ({\hyperref[fig:5]{Fig. 5}} (a)) and NLoS locations ({\hyperref[fig:5]{Fig. 5}} (b)), as described in the map in {\hyperref[fig:3]{Fig 3}}. In both cases, the average path loss values (both measured and simulated) have a trend which is in good agreement with the 3GPP InH reference model. It is worth noting that both RT tools show a good agreement with measurements. From simulations some insight about propagation mechanisms can be inferred: in the JMA office environment propagation is dominated by the LoS path (for the ``Meeting Room'') or low-loss NLoS paths transmitted through the light partition walls (for the ``Guest Room''). In other terms, in most NLoS locations of the considered office environment transmission through walls is the dominant propagation mechanism, as it will be shown in the following (see Section {\hyperref[sec:4]{IV}}.{\hyperref[sec:4C]{C}}), due to the small attenuation of plasterboard partition walls. Therefore, this kind of open-plan indoor environments are quite suitable for 27 and 38 GHz transmission despite the presence of partition walls, as confirmed by the high levels of received power generally observed.

\begin{figure}[!ht]
\centering
\includegraphics[width=0.47\textwidth]{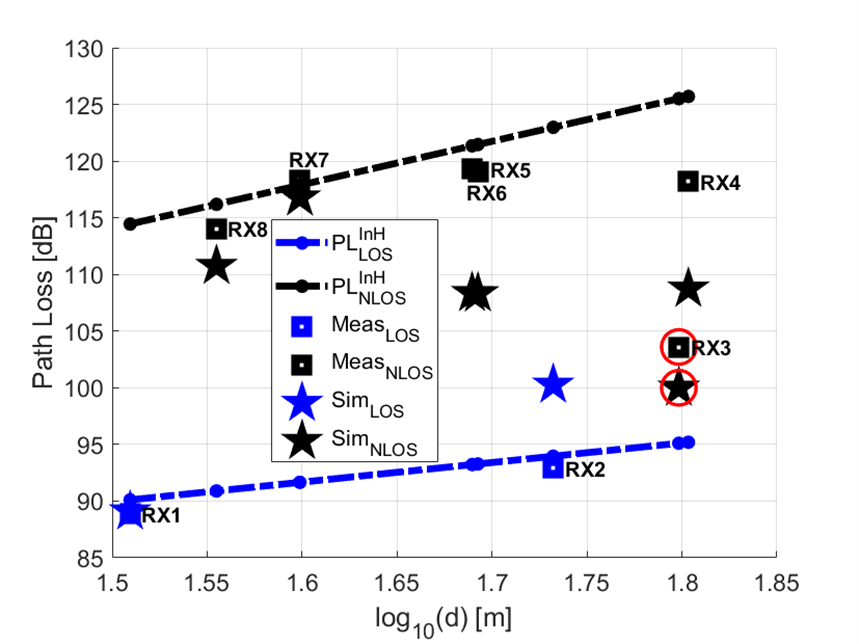}
\caption{Measured and simulated PL, and 3GPP UMi model, as a function of the distance, for the outdoor UniBO environment @ 38 GHz.}
\label{fig:6}
\end{figure}

In {\hyperref[fig:6]{Fig 6}} outdoor measurements at the University campus are compared with the Urban Micro (UMi) model. The UMi scenario is intended to represent street canyons and open areas, with the transmitter mounted below the rooftop levels of surrounding buildings (less than 10 m), and the receivers are at a height of 1.5-2.5 m.

Furthermore, also in this outdoor scenario there are some RXs
in a quasi-LoS condition, e.g. RX3 (highlighted by the red circle marker), where the 3GPP model doesn't perform well. The lower value of the PL with respect to the NLOS case, may be due to the presence of strong reflections on the metal advertising board panel located behind the RX3.

\subsection{Angular domain analysis}

\begin{figure}[!ht]
\centering
\includegraphics[width=0.45\textwidth]{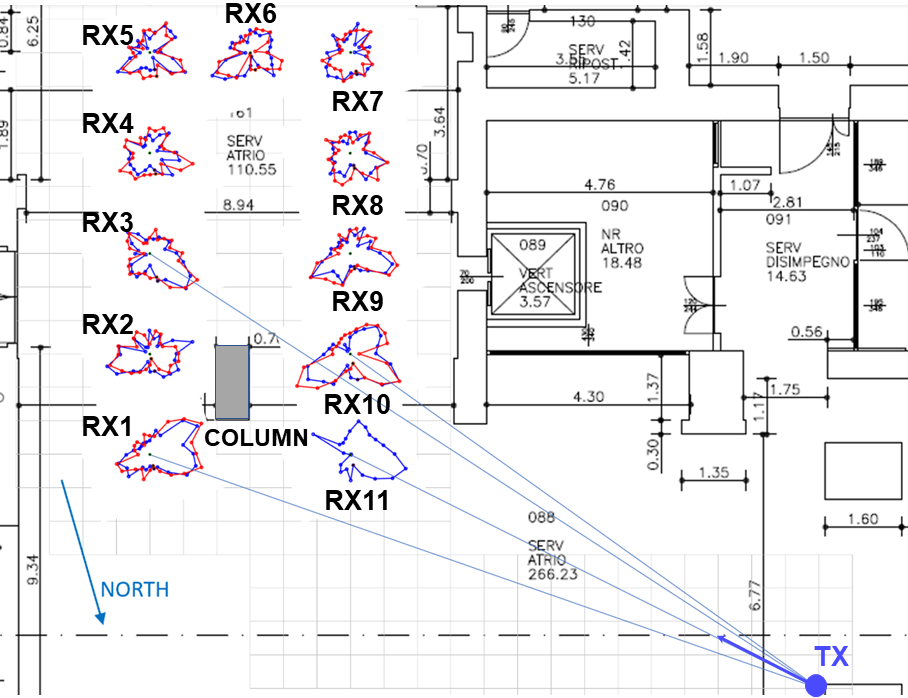}
\caption{Measured PAPs for the indoor UniBO scenario @ 27 GHz (blue line) and 38 GHz (red line).}
\label{fig:7}
\end{figure}

\begin{figure}[!ht]
\centering
\includegraphics[width=0.49\textwidth]{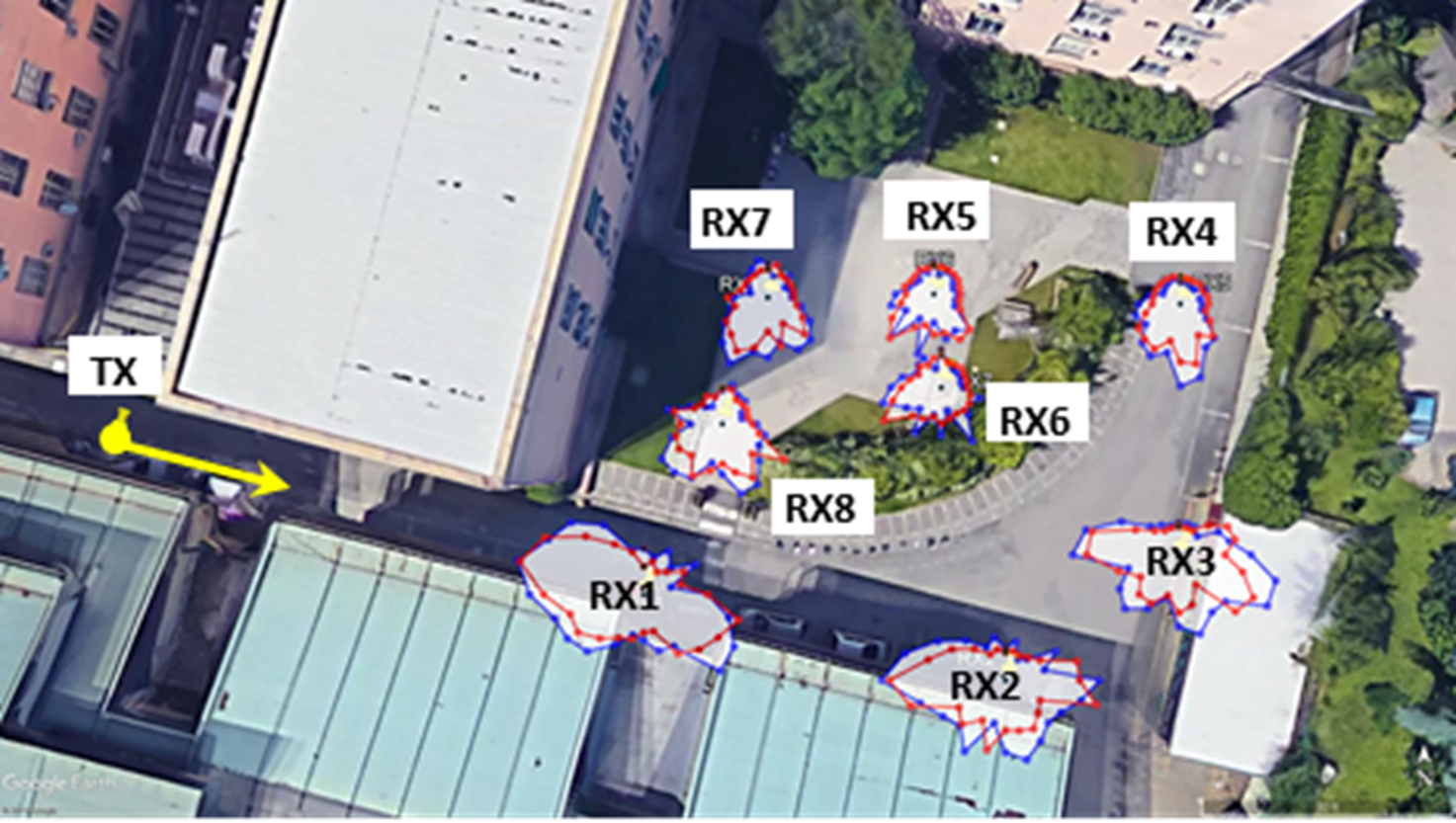}
\caption{Measured PAPs for the outdoor UniBO scenario @ 27 GHz (blue line) and 38 GHz (red line).}
\label{fig:8}
\end{figure}

In order to investigate the multipath spatial properties of the wireless channel, a directional survey is considered in both indoor and outdoor University scenarios (see {\hyperref[fig:1]{Fig 1}} and {\hyperref[fig:2]{Fig 2}}). Both the Power Angle Profile (PAP) related to the (azimuth) angle-of-arrival at the Rx side and the Angle Spread (AS) have been considered. Since the PAPs depend on the environment and on each Rx location, they have been plotted over the environment layout at each RX position, as shown in {\hyperref[fig:7]{Fig 7}} and {\hyperref[fig:8]{Fig 8}} for the indoor and outdoor case, respectively. PAPs at 27 and 38 GHz are depicted using blue and red colors, respectively. In {\hyperref[fig:7]{Fig 7}} the LoS path is highlighted with blue lines, where it is present.

\begin{figure}[!ht]
\centering
\subfloat[]{\includegraphics[width=0.24\textwidth]{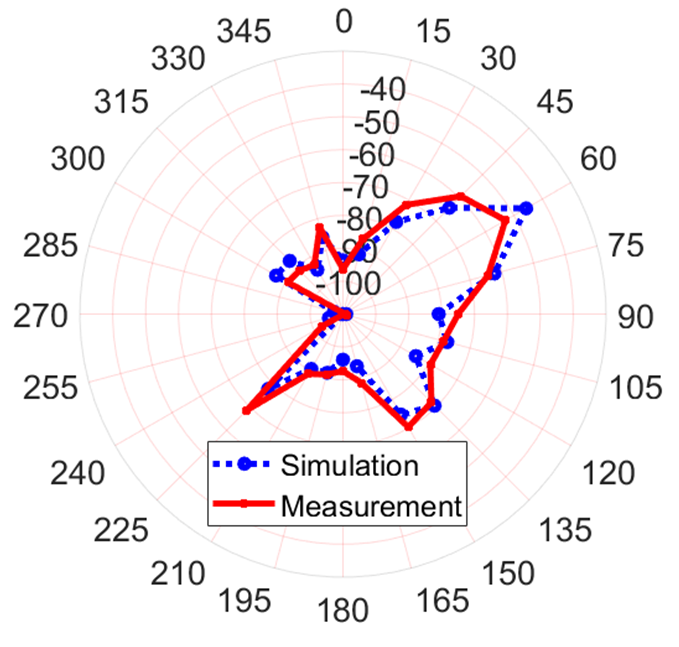}}
\hfill
\subfloat[]{\includegraphics[width=0.24\textwidth]{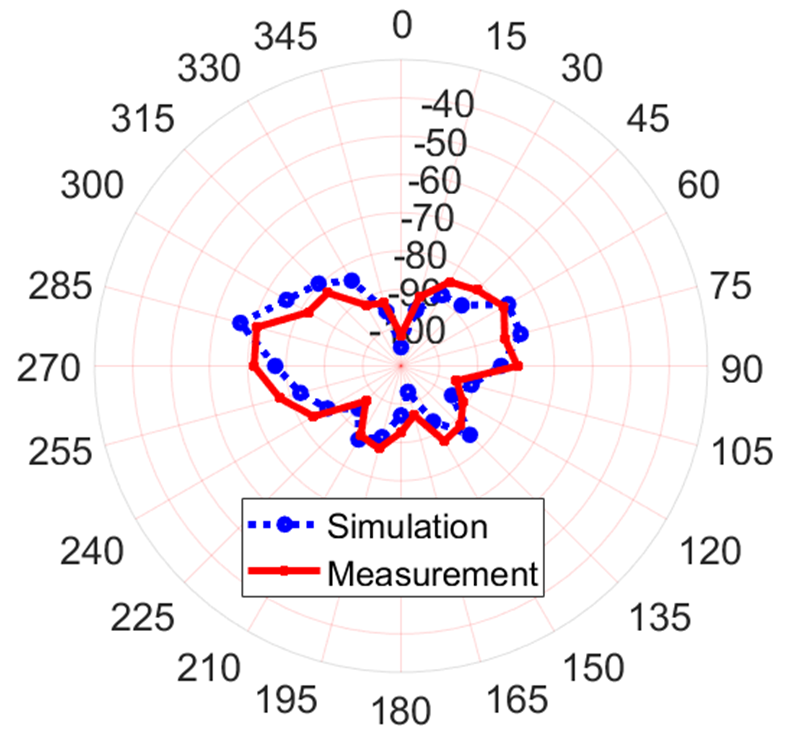}}
\caption{Comparison between measured and simulated PAP for the indoor UniBO scenario @ 27 GHz: (a) RX11 (LOS); (b) RX2 (NLOS).}
\label{fig:9}
\end{figure}

In {\hyperref[fig:9]{Fig 9}} the simulated (blue) and measured (red) PAP plots are shown for receivers RX11 an RX2 at 27 GHz in indoor scenario. The RMSE value is of 4.6 dB and 5.1 dB for receiver RX11 (LoS) and receiver RX2 (NLoS) respectively, which confirms the accuracy of the RT tool. From {\hyperref[fig:9]{Fig 9}} it is evident that the AS is larger in NLoS condition ({\hyperref[fig:9]{Fig 9}} (b)), than in LoS condition ({\hyperref[fig:9]{Fig 9}} (a)), as it should be due to the presence of the dominant LoS path in the latter case.

\begin{figure}[!ht]
\centering
\includegraphics[width=0.47\textwidth]{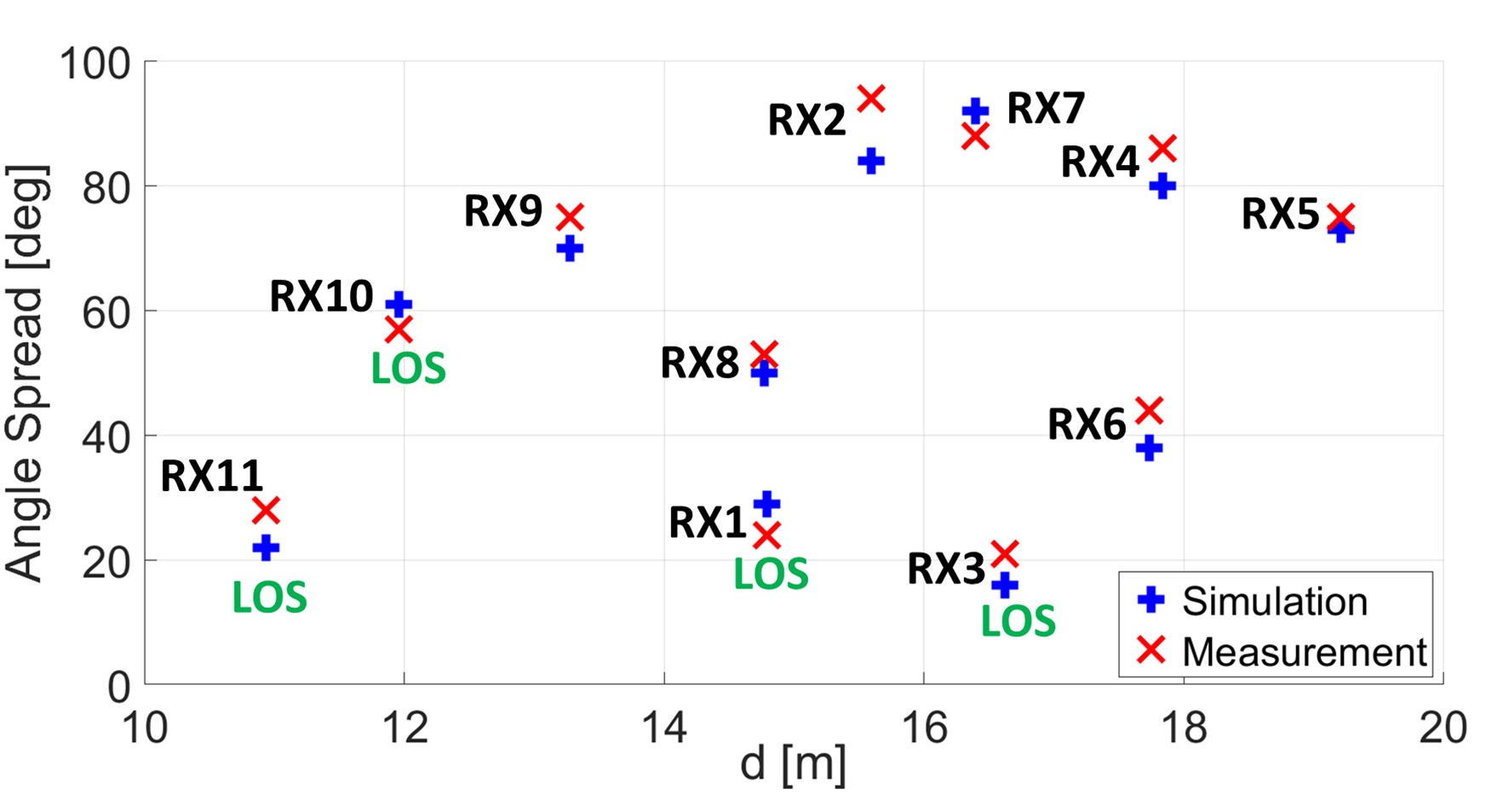}
\caption{Comparison between measured and simulated AS for the indoor UniBO scenario @ 27 GHz as a function of the TX-RX distance.}
\label{fig:10}
\end{figure}

This result is confirmed in {\hyperref[fig:10]{Fig 10}}, where the AS values are quite correlated to the LoS/NLoS condition. Conversely, they seem rather independent of the distance between TX and RXs.

RX10, which is in LoS, has a high AS, probably because in addition to the strong lobe in the LoS direction, there are also other two strong lobes coming from strong reflections from the central column and from the wall as reported in {\hyperref[fig:7]{Fig 7}}.

\begin{figure}[!ht]
\centering
\includegraphics[width=0.47\textwidth]{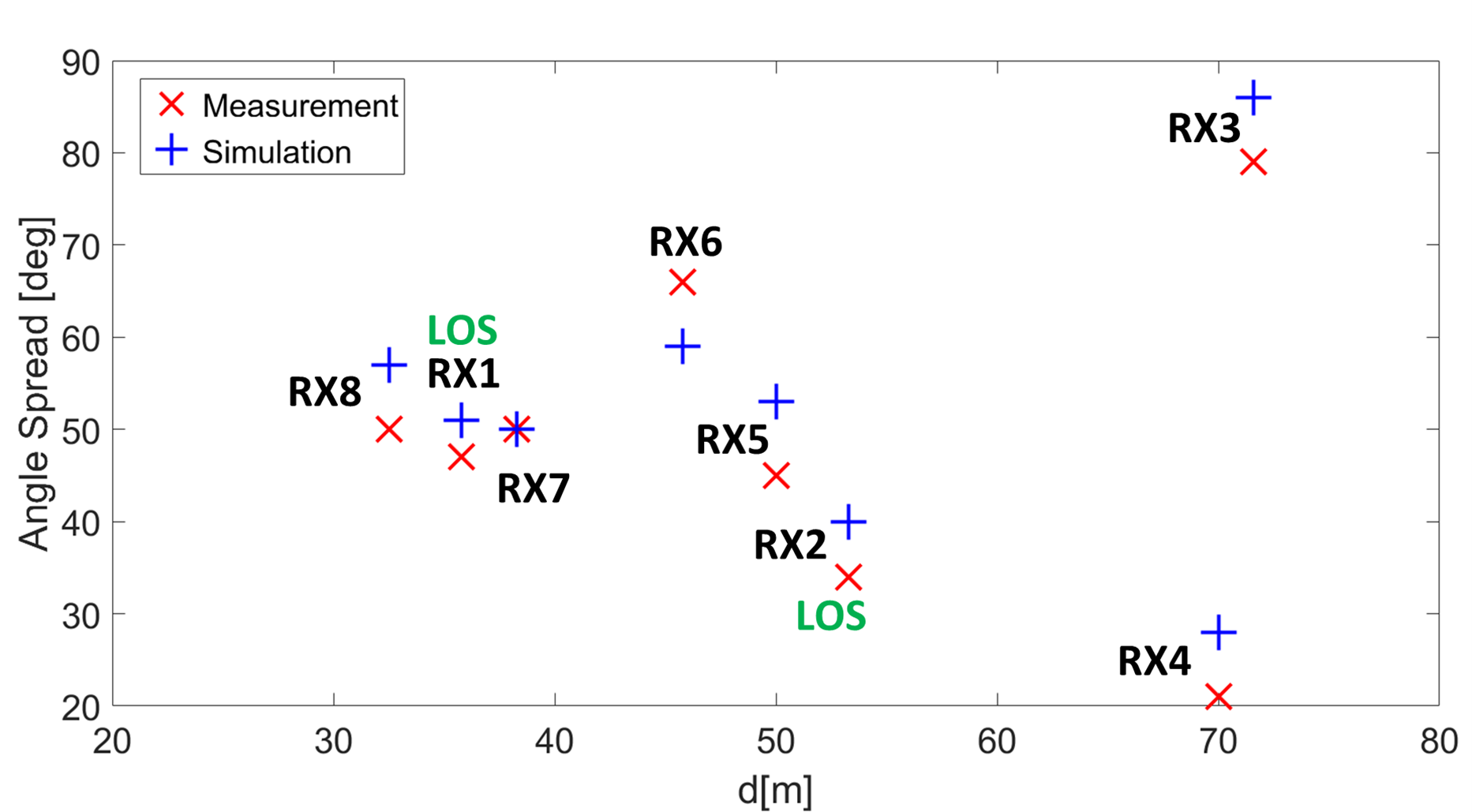}
\caption{Comparison between measured and simulated AS for the outdoor UniBO scenario @ 27 GHz as a function of the TX-RX distance}
\label{fig:11}
\end{figure}

With reference to the outdoor case, results reported in {\hyperref[fig:11]{Fig 11}}, confirm that angular dispersion is basically not affected by link distance. Differently from indoor, the AS looks weakly correlated also to the LoS/NLoS condition. As a matter of fact, the NLoS receiver RX4 shows a quite low AS, probably due to some strong signal contributions coming from the metal wall highlighted in {\hyperref[fig:2]{Fig 2}} and {\hyperref[fig:8]{Fig 8}}. At the same time, strong reflections are also likely to be present at RX1, thus increasing the AS in spite of the LoS condition. Although the lack of correlation between LoS/NLoS condition and AS is influenced by the characteristics of the investigated scenario, the result has general validity for the case of dense urban scenarios, since strong reflectors such as advertising panels or glass walls with metal-film are very common in urban environment.

In conclusion, strong multipath components seem to rise up in some indoor and outdoor locations, thus greatly affecting signal propagation in both LoS and NLoS conditions, increasing the angle spread in the former case but instead reducing angular dispersion in the latter.

\begin{figure}[!ht]
\centering
\includegraphics[width=0.49\textwidth]{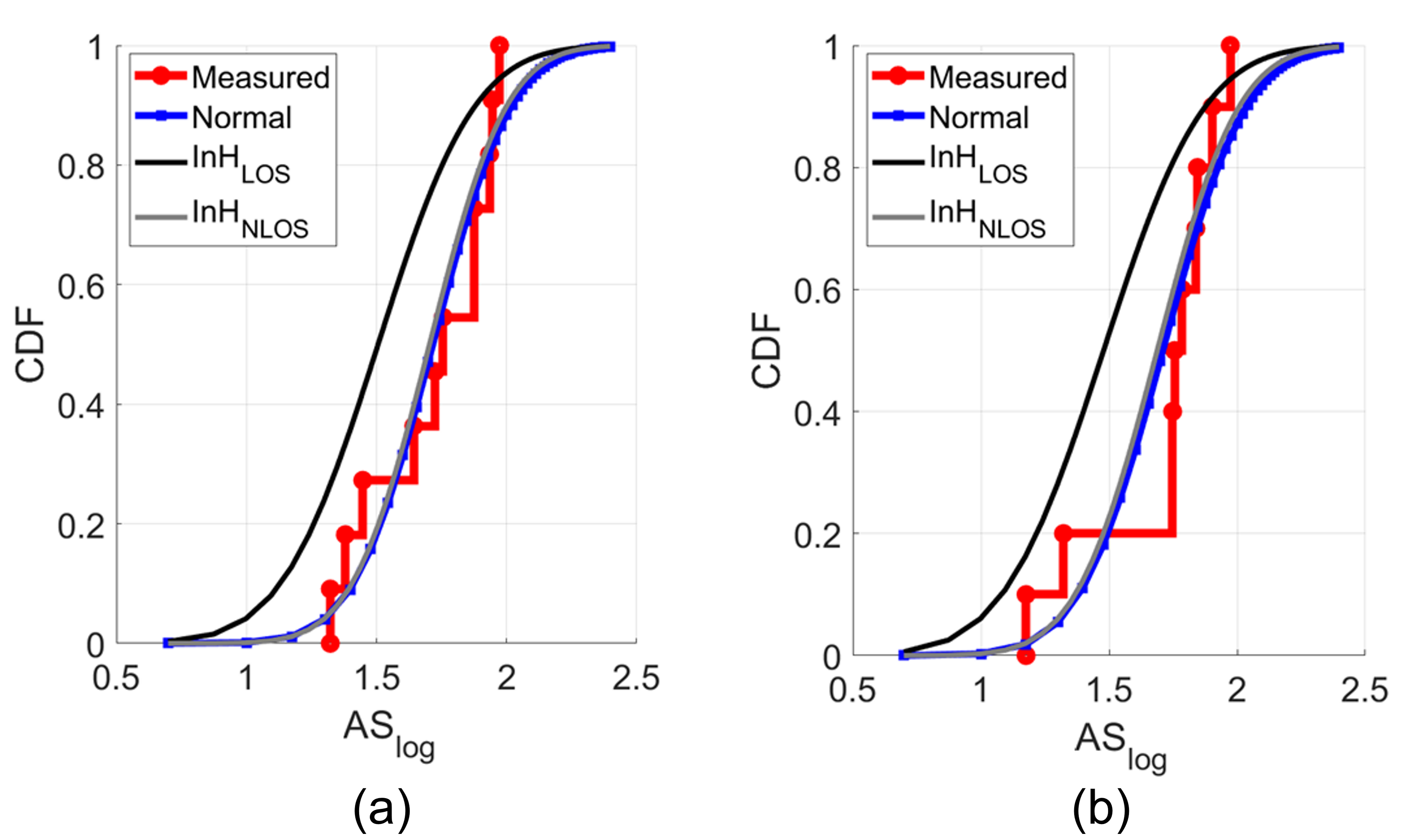}
\caption{Cumulative probability distribution function of AS in the indoor UniBO scenario @ 27 GHz (a) and 38 GHz (b).}
\label{fig:12}
\end{figure}

\begin{figure}[!ht]
\centering
\includegraphics[width=0.49\textwidth]{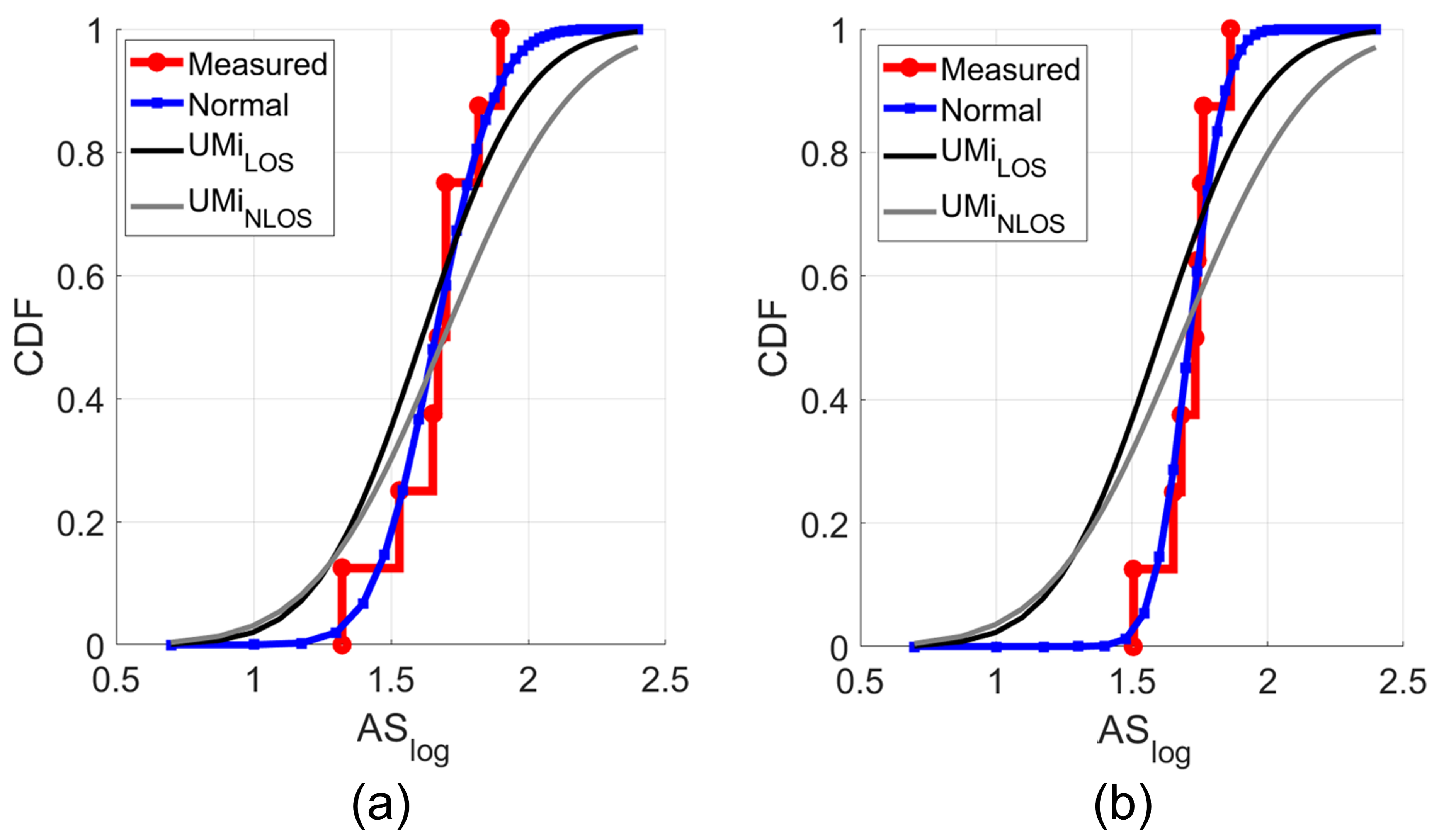}
\caption{Cumulative probability distribution function of AS around buildings (outdoor UniBO scenario) @ 27 GHz (a) and 38 GHz (b).}
\label{fig:13}
\end{figure}

{\hyperref[fig:12]{Fig 12}} and {\hyperref[fig:13]{Fig 13}} show the cumulative distribution functions (CDFs) plotted as a function of logarithmic AS, defined as \cite{22}: $AS_{log}=log_{10}(AS/1^{\circ})$, for the two considered frequencies in indoor and outdoor scenario respectively. The AS for indoor scenario are in the range of 21$^{\circ}$-94$^{\circ}$ for 27 GHz and 15$^{\circ}$-93$^{\circ}$ for 38 GHz, while the outdoor ASs are in the range of 21$^{\circ}$-79$^{\circ}$ for 27 GHz and 32$^{\circ}$-73$^{\circ}$ for 38 GHz.

All the measured AS values (in red) are fitted with a Gaussian Normal distribution N \textasciitilde{} (${\upmu}$, ${\upsigma^2}$), plotted in {\hyperref[fig:12]{Fig 12}} and {\hyperref[fig:13]{Fig 13}} with a blue line. The distribution parameters ${\upmu}$ and ${\upsigma}$ \textendash{} i.e. mean value and standard deviation \textendash{} are shown in Table IV and Table V for the indoor and outdoor scenario, respectively.

\begin{table}[!ht]
\caption{Gaussian distribution parameters of measured AS vs 3GPP AS model (indoor scenario)}
\begin{tabularx}{0.49\textwidth}{|
p{\dimexpr 0.18\linewidth-2\tabcolsep-2\arrayrulewidth}|
p{\dimexpr 0.15\linewidth-2\tabcolsep-\arrayrulewidth}|
p{\dimexpr 0.16\linewidth-2\tabcolsep-\arrayrulewidth}|
p{\dimexpr 0.16\linewidth-2\tabcolsep-\arrayrulewidth}|
p{\dimexpr 0.18\linewidth-2\tabcolsep-\arrayrulewidth}|
p{\dimexpr 0.17\linewidth-2\tabcolsep-\arrayrulewidth}|} \hline
\multicolumn{2}{|p{\dimexpr 0.33\linewidth-2\tabcolsep-2\arrayrulewidth}|}{\centering\arraybackslash{}\multirow[t]{2}{=}{\centering\arraybackslash{}} } & \multicolumn{2}{p{\dimexpr 0.32\linewidth-2\tabcolsep-\arrayrulewidth}|}{\centering\arraybackslash{}\centering\arraybackslash{}\textbf{3GPP MODEL}} & \multicolumn{2}{p{\dimexpr 0.35\linewidth-2\tabcolsep-\arrayrulewidth}|}{\centering\arraybackslash{}\centering\arraybackslash{}\textbf{MEASUREMENT}} \\\cline{3-6}
\multicolumn{2}{|l|}{} & \centering\arraybackslash{}$\boldsymbol{\upmu}$ & \centering\arraybackslash{}$\boldsymbol{\upsigma}$ & \centering\arraybackslash{}$\boldsymbol{\upmu}$ & \centering\arraybackslash{}$\boldsymbol{\upsigma}$ \\\hline
\centering\arraybackslash{}\multirow{2}{=}{\centering\arraybackslash{}\textbf{27 GHz}}  & \centering\arraybackslash{}LOS & \centering\arraybackslash{}1.5060 & \centering\arraybackslash{}0.2927 & \centering\arraybackslash{}\multirow{2}{=}{\centering\arraybackslash{}1.7159}  & \centering\arraybackslash{}\multirow{2}{=}{\centering\arraybackslash{}0.2374}  \\\cline{2-4}
 & \centering\arraybackslash{}NLOS & \centering\arraybackslash{}1.7038 & \centering\arraybackslash{}0.2327 &  &  \\\hline
\centering\arraybackslash{}\multirow{2}{=}{\centering\arraybackslash{}\textbf{38 GHz}}  & \centering\arraybackslash{}LOS & \centering\arraybackslash{}1.4787 & \centering\arraybackslash{}0.3099 & \centering\arraybackslash{}\multirow{2}{=}{\centering\arraybackslash{}1.7096}  & \centering\arraybackslash{}\multirow{2}{=}{\centering\arraybackslash{}0.2555}  \\\cline{2-4}
 & \centering\arraybackslash{}NLOS & \centering\arraybackslash{}1.6880 & \centering\arraybackslash{}0.2499 &  &  \\\hline
\end{tabularx}
\end{table}

\begin{table}[!ht]
\caption{Gaussian distribution parameters of measured AS vs 3GPP AS model (outdoor scenario)}
\begin{tabularx}{0.49\textwidth}{|
p{\dimexpr 0.18\linewidth-2\tabcolsep-2\arrayrulewidth}|
p{\dimexpr 0.15\linewidth-2\tabcolsep-\arrayrulewidth}|
p{\dimexpr 0.16\linewidth-2\tabcolsep-\arrayrulewidth}|
p{\dimexpr 0.16\linewidth-2\tabcolsep-\arrayrulewidth}|
p{\dimexpr 0.18\linewidth-2\tabcolsep-\arrayrulewidth}|
p{\dimexpr 0.17\linewidth-2\tabcolsep-\arrayrulewidth}|} \hline
\multicolumn{2}{|p{\dimexpr 0.33\linewidth-2\tabcolsep-2\arrayrulewidth}|}{\multirow[t]{2}{=}{} } & \multicolumn{2}{p{\dimexpr 0.32\linewidth-2\tabcolsep-\arrayrulewidth}|}{\centering\arraybackslash{}\textbf{3GPP MODEL}} & \multicolumn{2}{p{\dimexpr 0.35\linewidth-2\tabcolsep-\arrayrulewidth}|}{\centering\arraybackslash{}\textbf{MEASUREMENT}} \\\cline{3-6}
\multicolumn{2}{|l|}{} & \centering\arraybackslash{}$\boldsymbol{\upmu}$ & \centering\arraybackslash{}$\boldsymbol{\upsigma}$ & \centering\arraybackslash{}$\boldsymbol{\upmu}$ & \centering\arraybackslash{}$\boldsymbol{\upsigma}$ \\\hline
\multirow{2}{=}{\centering\arraybackslash{}\textbf{27 GHz}}  & \centering\arraybackslash{}LOS & \centering\arraybackslash{}1.6142 & \centering\arraybackslash{}0.3003 & \multirow{2}{=}{\centering\arraybackslash{}1.6618}  & \multirow{2}{=}{\centering\arraybackslash{}0.1756}  \\\cline{2-4}
 & \centering\arraybackslash{}NLOS & \centering\arraybackslash{}1.6942 & \centering\arraybackslash{}1.6827 &  &  \\\hline
\multirow{2}{=}{\centering\arraybackslash{}\textbf{38 GHz}}  & \centering\arraybackslash{}LOS & \centering\arraybackslash{}1.6027 & \centering\arraybackslash{}0.3023 & \multirow{2}{=}{\centering\arraybackslash{}1.7119}  & \multirow{2}{=}{\centering\arraybackslash{}0.1041}  \\\cline{2-4}
 & \centering\arraybackslash{}NLOS & \centering\arraybackslash{}1.6827 & \centering\arraybackslash{}0.3796 &  &  \\\hline
\end{tabularx}
\end{table}

In general, measurements at 38 GHz and/or outdoor locations tend to have smaller AS with respect to 27 GHz and/or indoor locations, highlighting a higher degree of multipath richness in outdoor environment and for lower frequencies.

In Fig.s {\hyperref[fig:12]{12}}\textendash{}{\hyperref[fig:13]{13}} and Tables IV-V the measured AS values are compared with the 3GPP models as defined in \cite{22}. The indoor measurements are compared with the indoor office scenario (InH) 3GPP model for both LoS and NLoS cases. From Table IV is clear that the NLoS case is in good agreement with the measured indoor data.

The outdoor scenario is compared with the Urban Micro (UMi) 3GPP model as defined in \cite{22}. The LoS case shows a better agreement with the measured data (see Table V), even though in the outdoor measurements there is a combination of LoS, quasi-LOS and NLoS RXs.

\subsection{Analysis of propagation mechanisms}\label{sec:4C}

Once validated in the considered environments, the RT tool has been used for evaluating the contribution of the different propagation mechanisms \textendash{} namely reflection, diffraction, diffuse scattering and combinations of them, as summarized in Table II \textendash{} to the total received power.

\begin{figure}[!ht]
\centering
\includegraphics[width=0.4\textwidth]{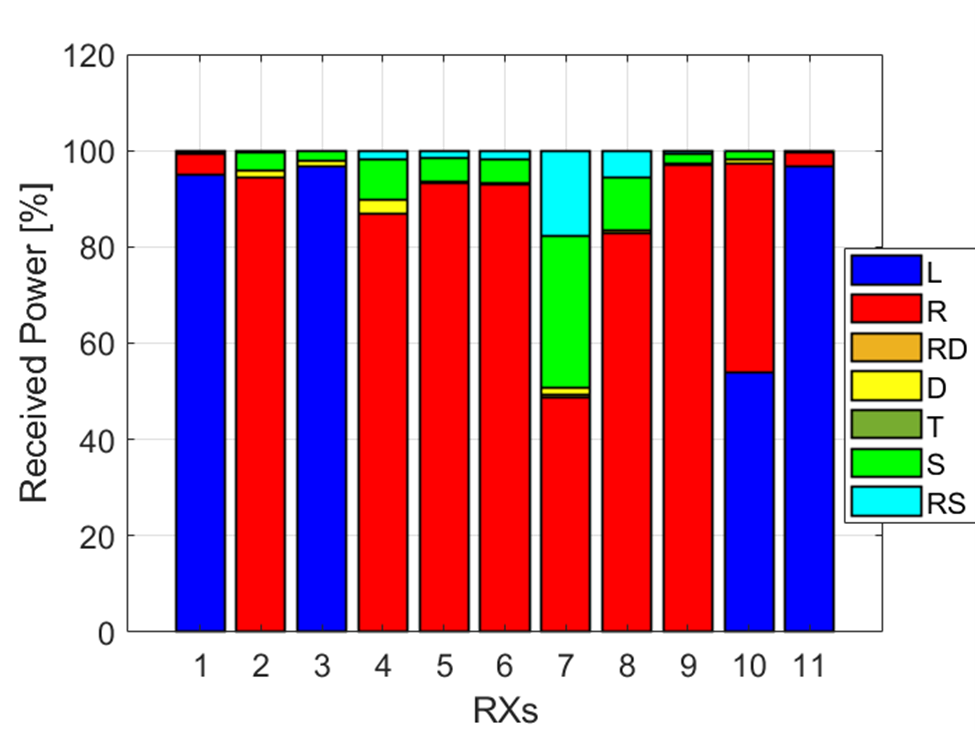}
\caption{Indoor scenario - Contribution of the different propagation mechanisms to the total received power at 27 GHz.}
\label{fig:14}
\end{figure}

\begin{figure}[!ht]
\centering
\includegraphics[width=0.4\textwidth]{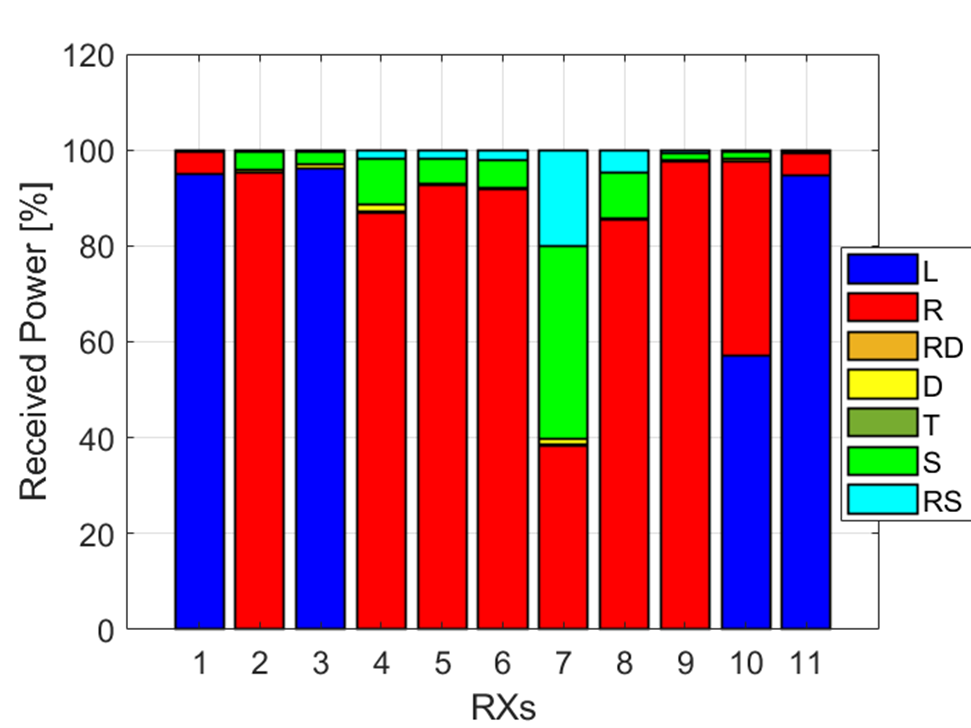}
\caption{Large indoor scenario - Contribution of the different propagation mechanisms to the total received power at 38 GHz.}
\label{fig:15}
\end{figure}

In {\hyperref[fig:14]{Fig 14}} and {\hyperref[fig:15]{Fig 15}} the simulation results for the large indoor scenario at 27 GHz and 38 GHz, respectively, are reported.

For LoS receivers (RX1, RX3, RX11 and RX10) and for both frequencies almost 100\% of the total received power is due to the direct ray (L), except for RX10, where a strong contribution from reflection (43\%) is observed, probably a single-bounce reflection coming from the central column (see {\hyperref[fig:1]{Fig 1}} and {\hyperref[fig:7]{Fig 7}}). Diffraction (D) is almost always negligible, as expected for mm-waves frequencies, as well as combined reflection and diffraction (RD) contributions, which are always below the 3\% of the total power.

\begin{figure}[!ht]
\centering
\subfloat[]{\includegraphics[width=0.24\textwidth]{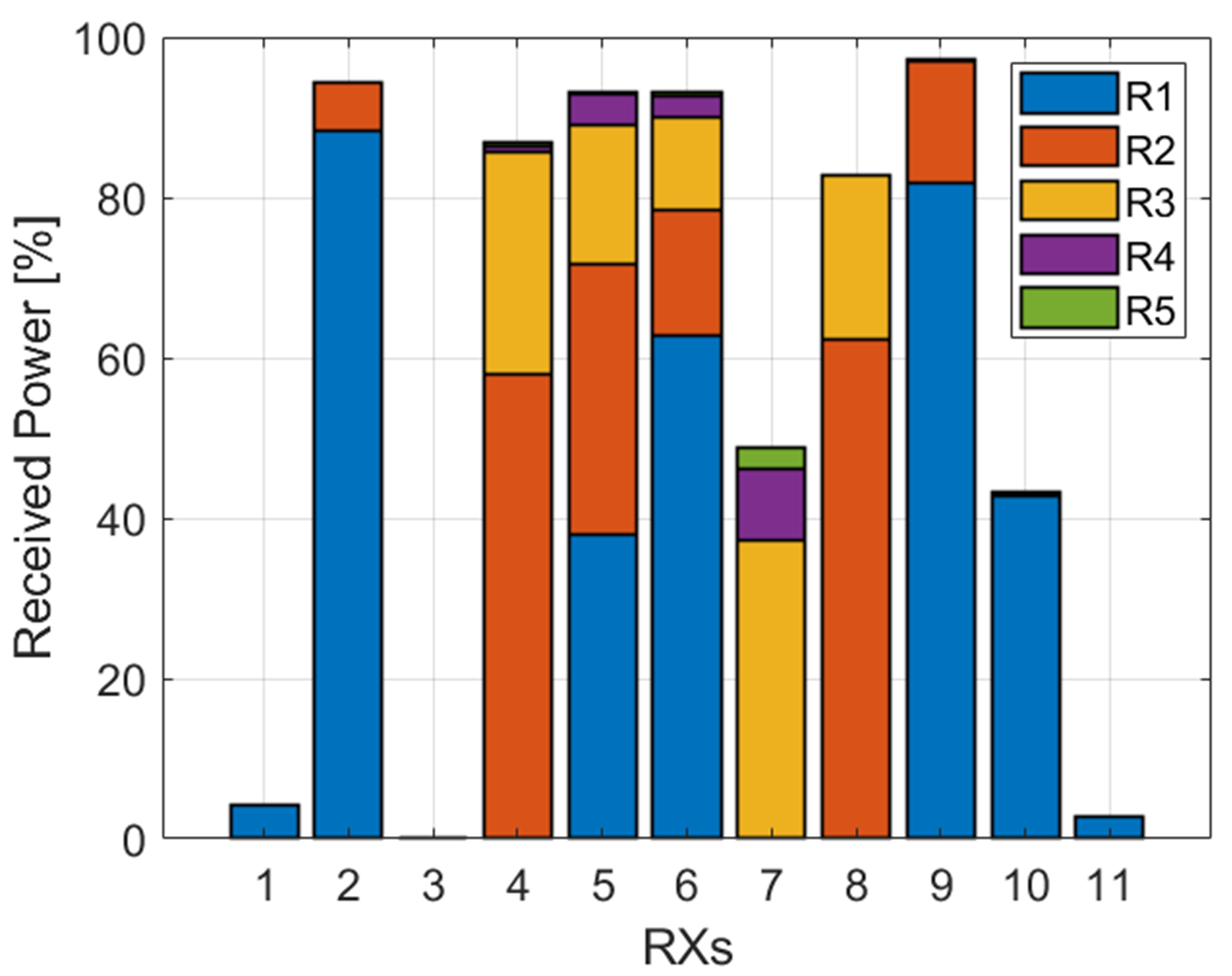}}
\hfill
\subfloat[]{\includegraphics[width=0.24\textwidth]{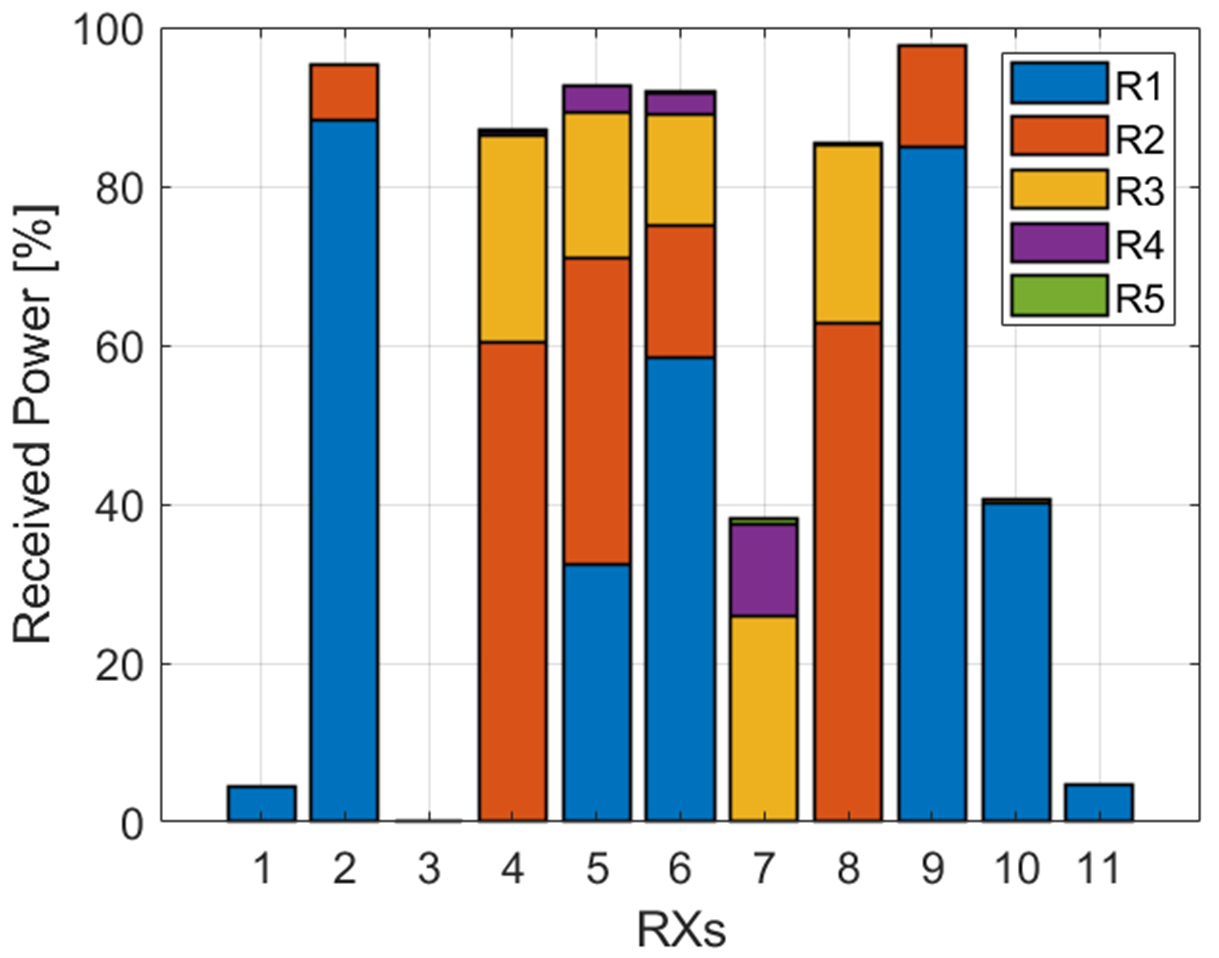}}
\caption{Insight of Reflection mechanism for 27 GHz (a) and 38 GHz (b) in the large indoor scenario.}
\label{fig:16}
\end{figure}

Reflection is the main propagation mechanism for NLoS receivers. As shown in {\hyperref[fig:16]{Fig 16}}, also high-order reflections up to the fourth-order are important for strongly NLoS receivers, such as RX7.

Furthermore, the contribution of diffuse scattering is fundamental for strongly NLoS receivers, such as RX7. In this case the 30\% and 40\% of the total received power @ 27 GHz and 38 GHz respectively is due to scattering, probably because of the presence of a few, big pieces of furniture not taken into account in RT simulation. As reported in section III, the ER diffuse scattering model implemented in the RT tool accounts for the surface roughness of the walls, decorations or smaller objects not present in the environment database. As shown in Table II, combined mechanisms have also been considered, such as reflections combined with scattering (RS). The total received power for NLoS receivers, at both frequencies appears to be due to reflections, scattering and a combination of reflections and scattering.

\begin{figure}[!ht]
\centering
\includegraphics[width=0.4\textwidth]{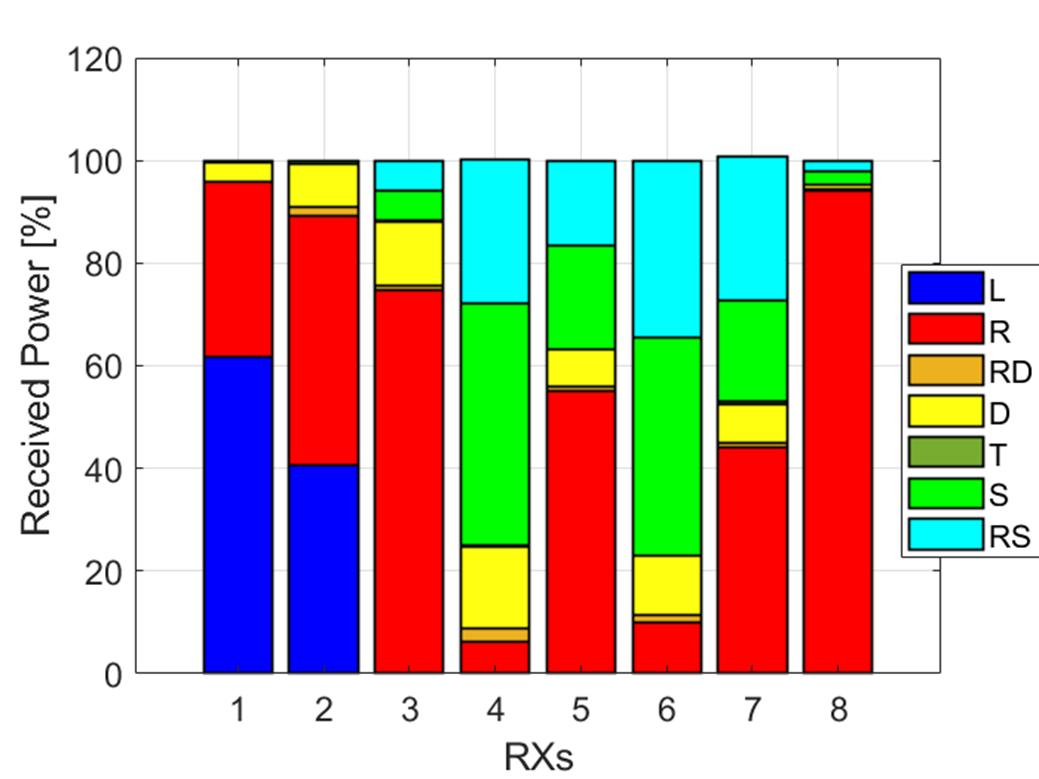}
\caption{Outdoor scenario - Contribution of the different propagation mechanisms total received power at 27 GHz.}
\label{fig:17}
\end{figure}

\begin{figure}[!ht]
\centering
\includegraphics[width=0.4\textwidth]{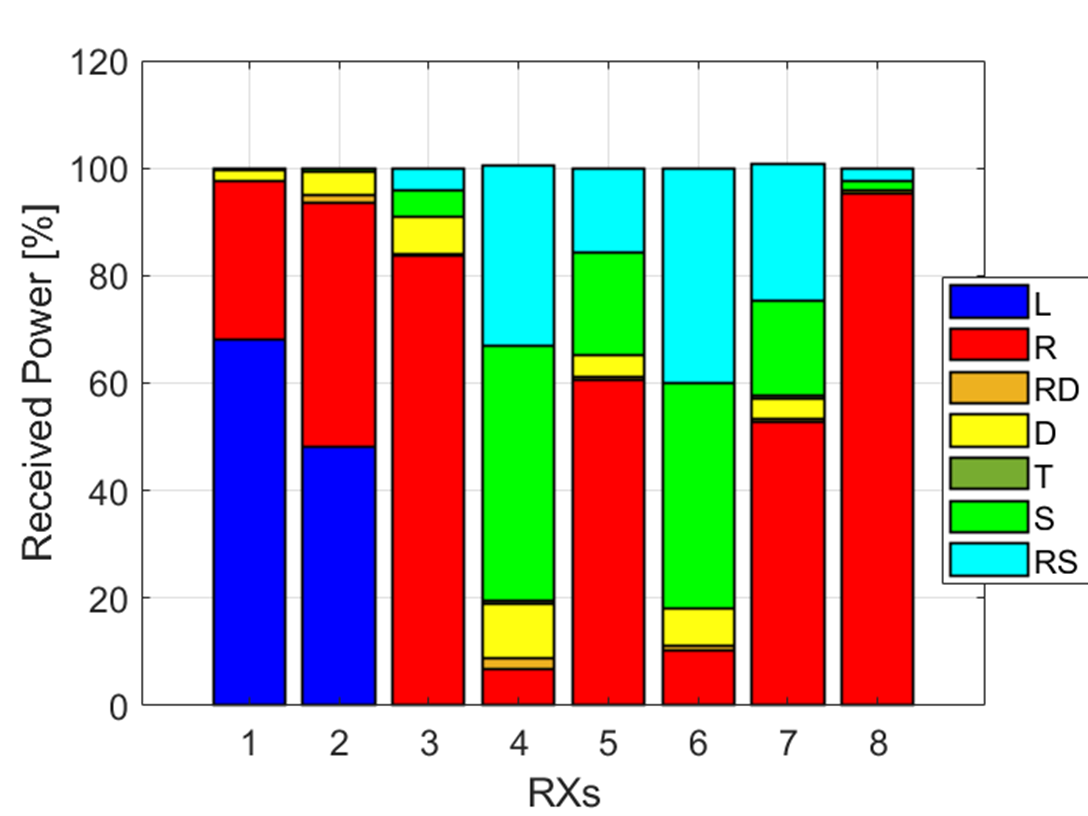}
\caption{Outdoor scenario - Contribution of the different propagation mechanisms total received power at 38 GHz.}
\label{fig:18}
\end{figure}

For what concerns the outdoor environment, {\hyperref[fig:17]{Fig 17}} and {\hyperref[fig:18]{Fig 18}} show simulation results at 27 GHz and 38 GHz respectively.

In the outdoor case the LoS contribution to the total received power is smaller than the indoor scenario, due to the presence of a higher multipath richness: the direct ray component for RX1 and RX2 is 67\% and 48\% respectively. However, it has to be considered that the TX-RXs distances are higher in this case: 35.8 m and 53.3 m respectively, while in the indoor case the farthest RX is at 16 m.

Specular reflection is significant also in LoS RXs, being greater than 35\% for RX1 and almost 50\% for RX2 at both frequencies. For NLoS RXs, such as RX3, RX5 and RX8 (see {\hyperref[fig:17]{Fig 17}} and {\hyperref[fig:18]{Fig 18}}), reflection is the main propagation mechanism.

\begin{figure}[!ht]
\centering
\subfloat[]{\includegraphics[width=0.24\textwidth]{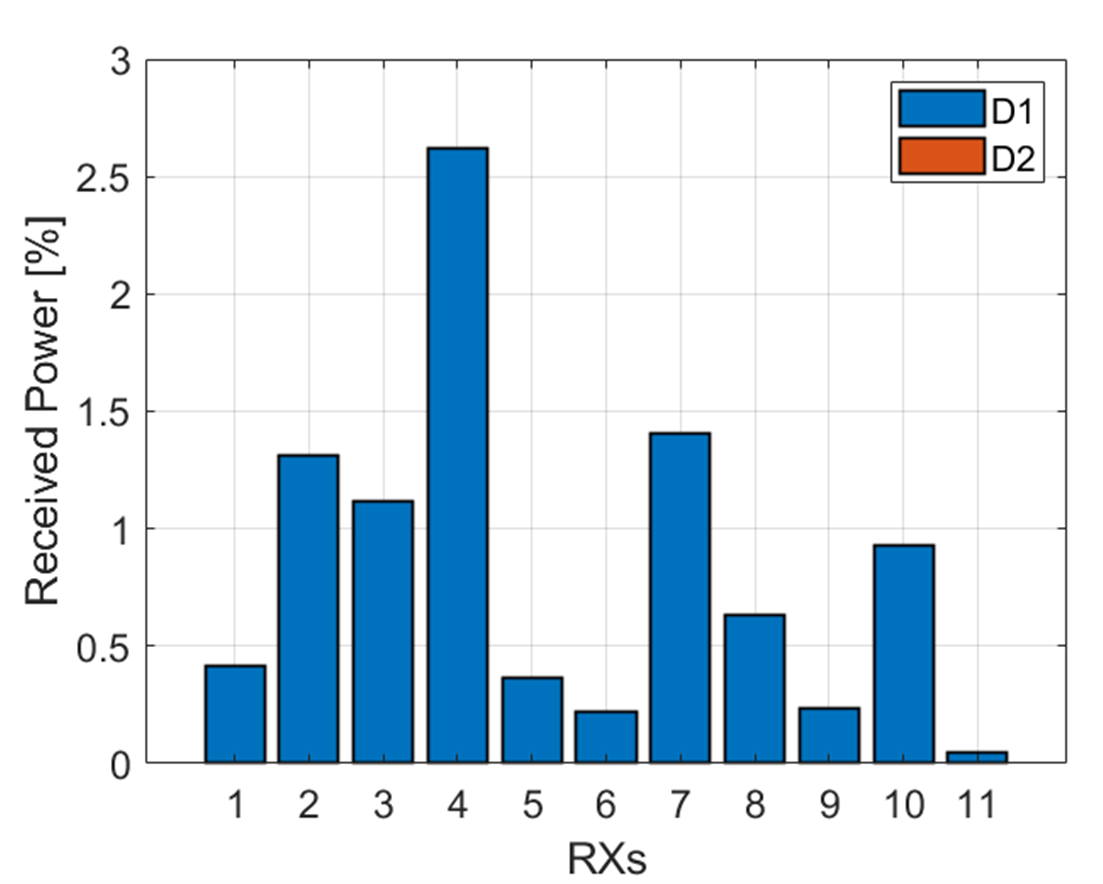}}
\hfill
\subfloat[]{\includegraphics[width=0.24\textwidth]{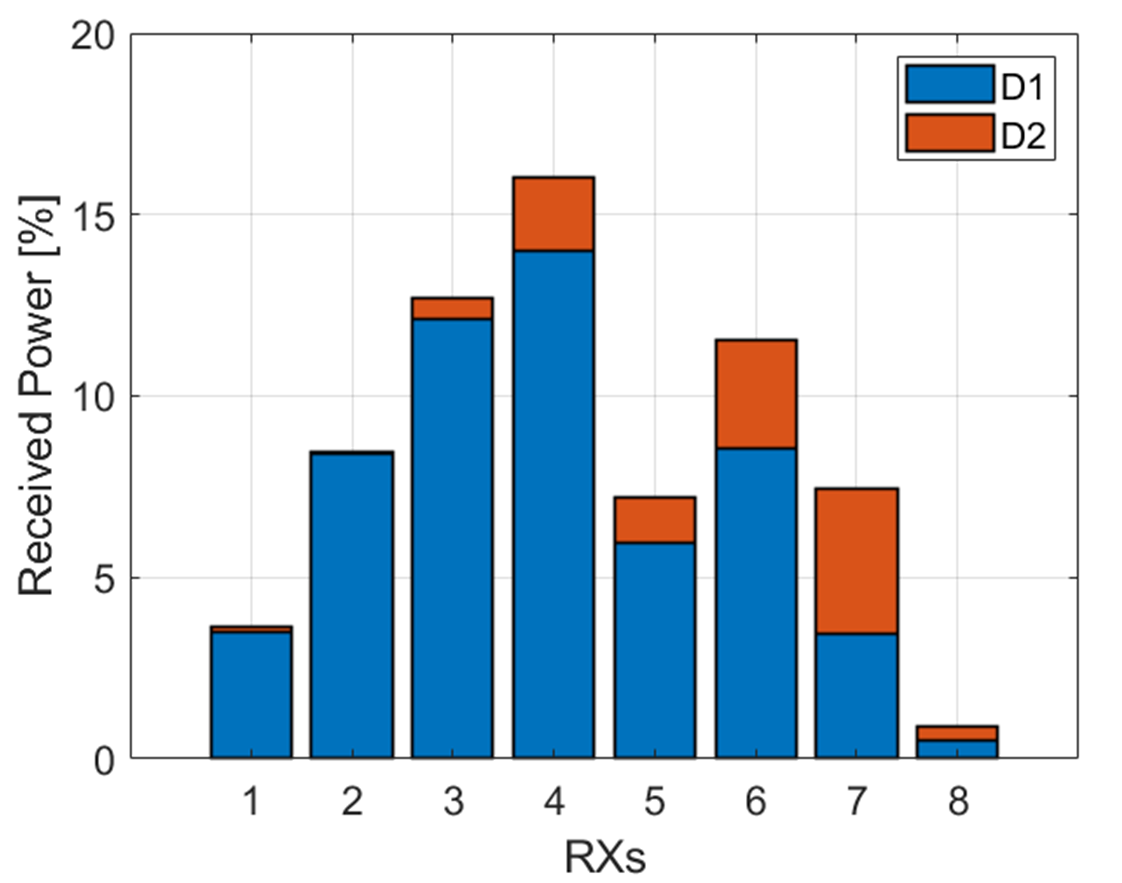}}
\caption{Insight of Diffraction mechanism a 27 GHz for indoor (left) and outdoor (right) scenarios.}
\label{fig:19}
\end{figure}

For NLoS RXs placed in the middle of the yard (RX4 and RX6), scattering alone and reflections combined with scattering are relevant, more than in the indoor case, probably due to the presence of bushes and bicycle stands. Moreover, the diffraction contribution is surprisingly higher in this case. In fact, while in the indoor scenario diffraction is always lower the 3\%, for the outdoor case diffraction reaches 10-15 \%, and even the second-order diffraction is detectable. As an example, {\hyperref[fig:19]{Fig 19}} reports the contribution of diffraction in indoor and outdoor scenarios at 27 GHz.

Reflections combined with diffraction and transmissions through walls are in any case negligible.

\begin{figure}[!ht]
\centering
\subfloat[]{\includegraphics[width=0.49\textwidth]{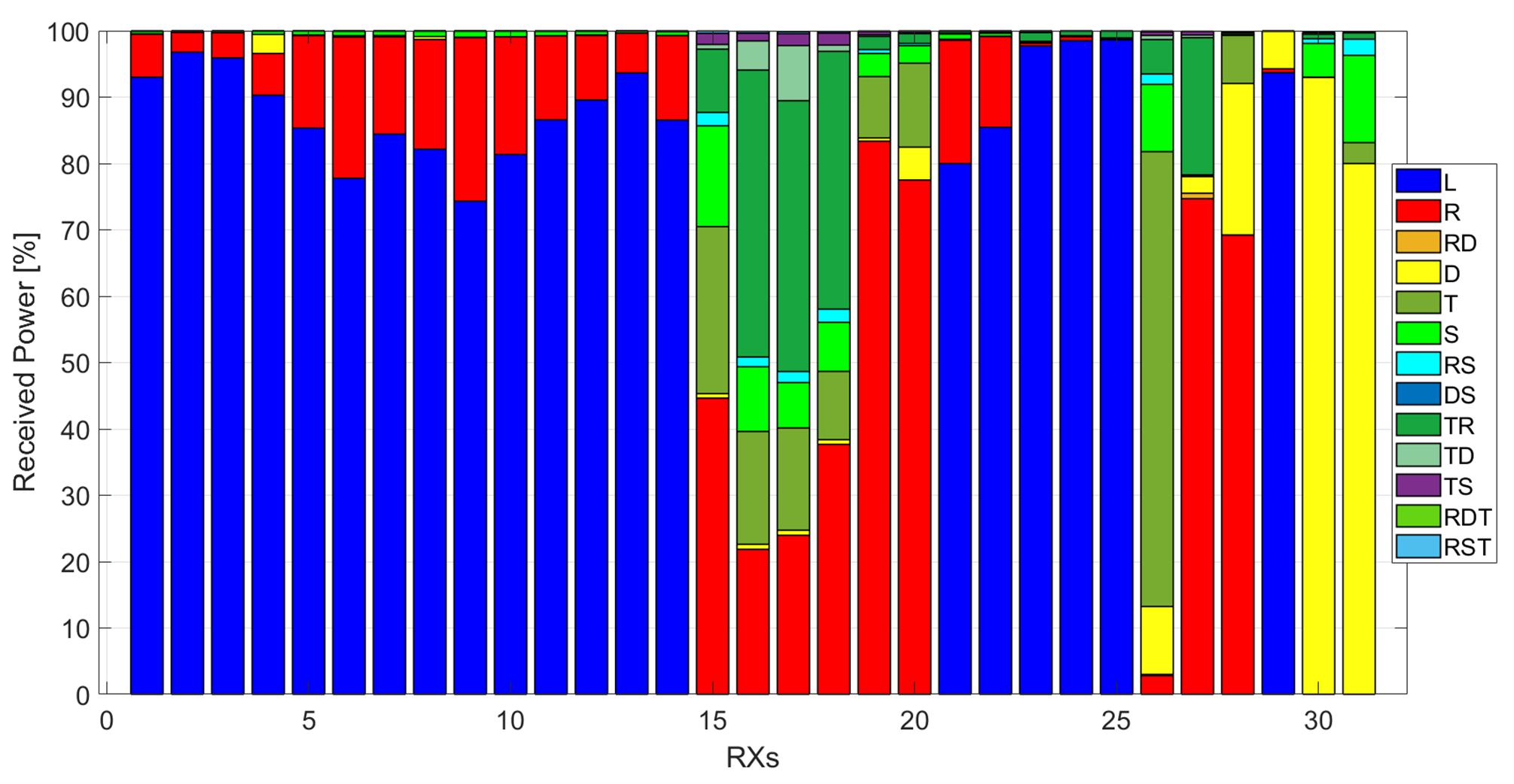}}
\hfill
\subfloat[]{\includegraphics[width=0.49\textwidth]{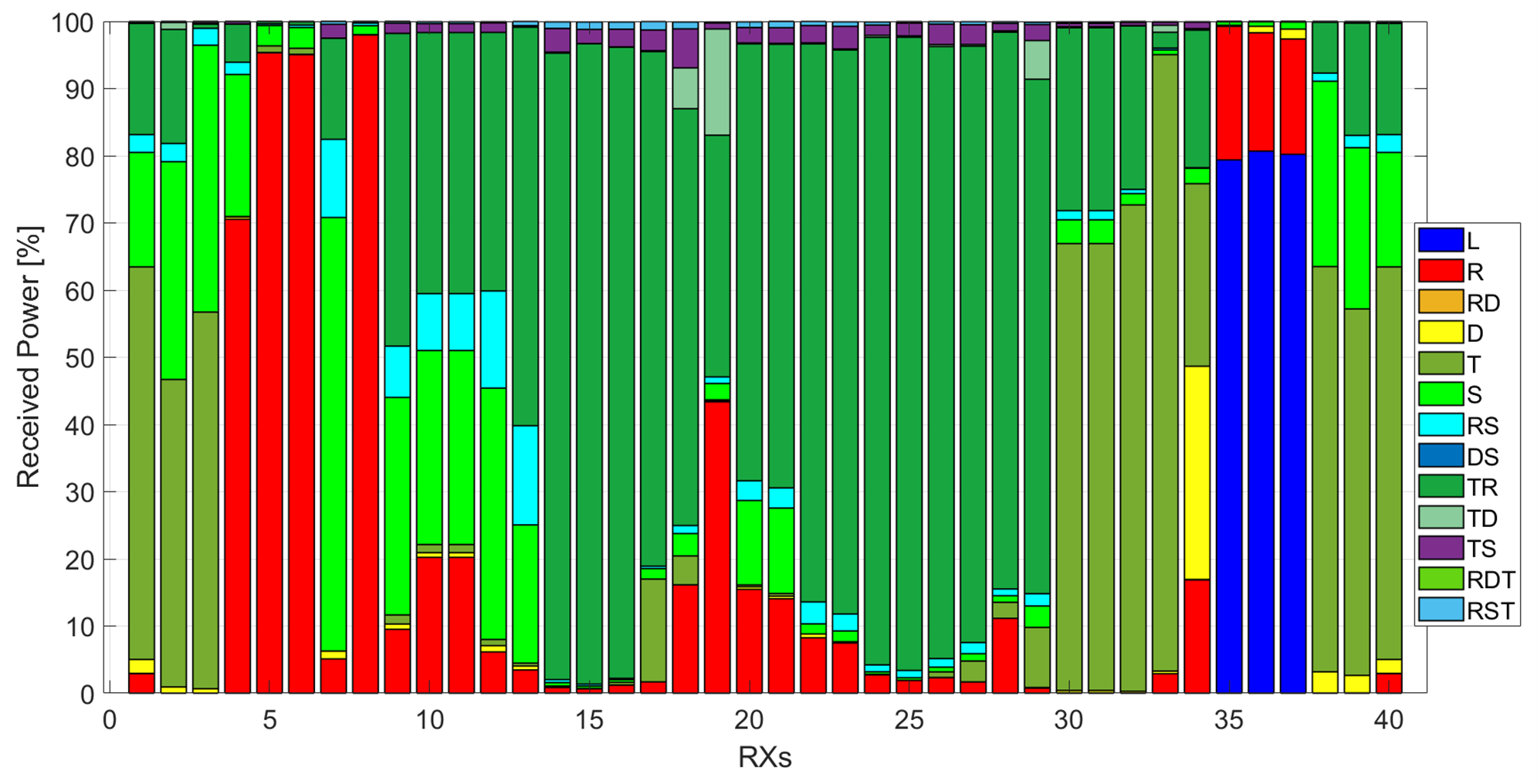}}
\caption{Contribution of the different propagation mechanisms in the indoor JMA scenario at 28 GHz, for both the internal (a) and external (b) receivers.}
\label{fig:20}
\end{figure}

Finally, in {\hyperref[fig:20]{Fig 20}} the analysis of the propagation mechanisms is shown for the indoor office scenario of JMA premises, including both the ``internal'' and ``external'' receiver routes (subfigures (a) and (b), respectively).

Interestingly, in most NLoS locations transmission (alone or combined with reflections) is largely dominant with respect to the other mechanisms, due to peculiar characteristics of the environment, which is essentially an open space with plasterboard partition walls.

Moreover, as already pointed out, reflections of order higher than 2 do not appear to be relevant because of the partition wall characteristics. As in the large indoor environment considered above, scattering, including combination with reflections appears to be significant in all NLoS locations, especially the external ones.

It’s also worth noting that in only in a few, very specific cases (e.g. locations 30-31 of {\hyperref[fig:20]{Fig 20}} (a)) diffraction can be the dominant mechanism: this usually happens in strongly NLoS locations as in this case, where RX is located on a corridor separated from the adjacent rooms through a thick concrete wall with high penetration loss, and it can be reached only through diffraction on the entrance door opening.

\section{Conclusion}\label{sec:5}
The results of an indoor and outdoor measurement and simulation campaign is presented in the paper, with the aim of investigating the characteristics of mm-wave propagation and provide insight for a reliable implementation of advanced transmission techniques.

Results confirm the much smaller contribution of diffraction with respect to sub-6 GHz frequencies, which leads to large prediction errors in NLoS locations if multiple reflections \textendash{} up to the fourth order \textendash{} and diffuse scattering are neglected.

The analysis of the multipath spatial properties of the wireless channel has shown that the direct path \textendash{} if present \textendash{} usually carries most of the power, but strong multipath components can be somewhere present, reducing angular dispersion in line of sight and instead increasing it under non line of sight conditions.
Furthermore, differences between indoor and outdoor AS values are highlighted as function of the frequency (38 GHz shows lower AS with respect to 27 GHz).

Moreover, the analysis in the indoor office environment has shown that the dominant NLoS propagation mechanism is unexpectedly transmission at both frequencies, but this behavior is due to the light-material of indoor-partition walls, that allow for good coverage also in NLoS conditions.

Follow up activities will include the analysis of propagation mechanisms in other typical environments, such as urban street canyons or small indoor offices, to validate the findings of the present work and confirm the use of deterministic ray-based propagation models as useful analysis and simulation tools.

% or


%\appendices

%\section{}

%\section*{Acknowledgment}

\ifCLASSOPTIONcaptionsoff
  \newpage
\fi



%\bibliographystyle{IEEEtran}

\begin{thebibliography}{1}

\bibitem{1} M. Shafi et al., "5G: A Tutorial Overview of Standards, Trials, Challenges, Deployment, and Practice," in \textit{IEEE J. Sel. Areas Commun.}, vol. 35, no. 6, pp. 1201-1221, June 2017, doi: 10.1109/JSAC.2017.2692307.

\bibitem{2} H. Tataria, M. Shafi, A. F. Molisch, M. Dohler, H. Sj\"{o}land and F. Tufvesson, "6G Wireless Systems: Vision, Requirements, Challenges, Insights, and Opportunities," \textit{Proc. IEEE}, vol. 109, no. 7, pp. 1166-1199, July 2021, doi: 10.1109/JPROC.2021.3061701.

\bibitem{3} T. S. Rappaport et al., "Wireless Communications and Applications Above 100 GHz: Opportunities and Challenges for 6G and Beyond," \textit{IEEE Access}, vol. 7, pp. 78729-78757, 2019, doi: 10.1109/ACCESS.2019.2921522.

\bibitem{4} D. Gesbert, M. Kountouris, R. W. Heath Jr., C. Chae and T. Salzer, "Shifting the MIMO paradigm", \textit{IEEE Signal Process. Mag.}, vol. 24, no. 5, pp. 36-46, Sep. 2007.

\bibitem{5} M. Giordani, M. Polese, A. Roy, D. Castor and M. Zorzi, "A tutorial on beam management for 3GPP NR at mmWave frequencies", \textit{IEEE Commun. Surv. Tutor.}, vol. 21, no. 1, pp. 173-196, 1st Quart. 2019.

\bibitem{6} X. Zhang, G. Qiu, J. Zhang, L. Tian, P. Tang and T. Jiang, "Analysis of millimeter-wave channel characteristics based on channel measurements in indoor environments at 39 GHz", in Proc. 11th Int. Conf. Wireless Commun. Signal Process. (WCSP), pp. 1-6, Oct. 2019.

\bibitem{7} D. Chizhik, J. Du, R. Feick, M. Rodriguez, G. Castro and R. A. Valenzuela, "Path Loss and Directional Gain Measurements at 28 GHz for Non-Line-of-Sight Coverage of Indoors With Corridors," \textit{IEEE Trans. Antennas Propag.}, vol. 68, no. 6, pp. 4820-4830, June 2020, doi: 10.1109/TAP.2020.2972609.

\bibitem{8} K. Haneda, J. Jarvelainen, A. Karttunen, M. Kyro and J. Putkonen, "A statistical spatio-temporal radio channel model for large indoor environments at 60 and 70 GHz", \textit{IEEE Trans. Antennas Propag.}, vol. 63, no. 6, pp. 2694-2704, Jun. 2015.

\bibitem{9} K. Wangchuk, K. Umeki, T. Iwata, P. Hanpinitsak, M. Kim, K. Saito, et al., "Double directional millimeter wave propagation channel measurement and polarimetric cluster properties in outdoor urban pico-cell environment", \textit{IEICE Trans. Commun.}, vol. E100.B, pp. 1133-1144, 2017.

\bibitem{10} H. Ding et al., "Ray-tracing Based Channel Clustering and Analysis at 28 GHz in Conference Environment," in Proc. 2020 14th European Conference on Antennas and Propagation (EuCAP), 2020, pp. 1-5, doi: 10.23919/EuCAP48036.2020.9135078.

\bibitem{11} M. Kim, T. Iwata, K. Umeki, K. Wangchuk, J.-I. Takada and S. Sasaki, "Simulation based mm-wave channel model for outdoor open area access scenarios," in Proc. 2016 URSI Asia-Pacific Radio Science Conference (URSI AP-RASC), 2016, pp. 1292-1295, doi: 10.1109/URSIAP-RASC.2016.7601164.

\bibitem{12} J. J\"{a}rvel\"{a}inen, K. Haneda and A. Karttunen, "Indoor Propagation Channel Simulations at 60 GHz Using Point Cloud Data," \textit{IEEE Trans. Antennas Propag.}, vol. 64, no. 10, pp. 4457-4467, Oct. 2016, doi: 10.1109/TAP.2016.2598200.

\bibitem{13} I. Rodriguez et al., "Analysis of 38 GHz mmWave Propagation Characteristics of Urban Scenarios," in Proc. 21th European Wireless Conference, 2015, pp. 1-8.

\bibitem{14} SAF Tehnika - Microwave Radio Experts," SAF Tehnika, 2021. [Online]. Available: \url{https://www.saftehnika.com/en/spectrumanalyzer}.

\bibitem{15} E. M. Vitucci et al., "Tuning ray tracing for mm-wave coverage prediction in outdoor urban scenarios," \textit{Radio Science}, vol. 54, no. 11, pp. 1112-1128, Nov. 2019, doi: 10.1029/2019RS006869.

\bibitem{16} E. M. Vitucci, V. Degli-Esposti, F. Fuschini, J. S. Lu, M. Barbiroli, J. N. Wu, M. Zoli, J. Zhu, H.L. Bertoni, ``Ray Tracing RF Field Prediction: An Unforgiving Validation'', \text{Int. Journal Antennas Propag.}, 2015, Vol. 2015, n. 184608.

\bibitem{17} J. R. Abel and J. W. Wallace, "4-40 GHz Permittivity Measurements of Indoor Building Materials," in Proc. 2019 IEEE International Symposium on Antennas and Propagation and USNC-URSI Radio Science Meeting, 2019, pp. 105-106, doi: 10.1109/APUSNCURSINRSM.2019.8888911.

\bibitem{18} D. Ferreira, I. Cui\~{n}as, R. F. S. Caldeirinha and T. R. Fernandes, "A review on the electromagnetic characterisation of building materials at micro- and millimetre wave frequencies," in Proc. 8th European Conference on Antennas and Propagation (EuCAP 2014), 2014, pp. 145-149, doi: 10.1109/EuCAP.2014.6901713.

\bibitem{19} L. Possenti et al., "Improved Fabry-P\'{e}rot electromagnetic material characterization: Application and results," \textit{Radio Science}, vol. 55, no. 11, pp. 1-15, Nov. 2020, doi: 10.1029/2020RS007164.

\bibitem{20} F. Fuschini, M. Zoli, E. M. Vitucci, M. Barbiroli and V. Degli-Esposti, "A Study on Millimeter-Wave Multiuser Directional Beamforming Based on Measurements and Ray Tracing Simulations," \textit{IEEE Trans. Antennas Propag.}, vol. 67, no. 4, pp. 2633-2644, April 2019, doi: 10.1109/TAP.2019.2894271.

\bibitem{21} https://ibwave.com/

\bibitem{22} Report ITU-R M.2412-0 ``Guidelines for evaluation of radio interface technologies for IMT-2020''

\end{thebibliography}

\end{document}